\documentclass[default]{aastex702}
\makeatletter
\let\frontmatter@title@above=\relax
\makeatother
\usepackage{lipsum}

\usepackage{amsthm}
\usepackage{mathrsfs}
\usepackage{bbold}

\usepackage{array}
\usepackage{multirow}
\usepackage{booktabs}

\usepackage[dvipsnames]{xcolor}
\usepackage{placeins}
\usepackage{comment}
\usepackage[normalem]{ulem}

\usepackage{algorithm}
\usepackage{algpseudocode}
\usepackage{listings}

\usepackage{enumitem}
\usepackage{acronym}

\setdescription{leftmargin=8pt}
\usepackage{amsmath}
\journalinfo{your text here}

\hypersetup{
    colorlinks=true,
    linkcolor=blue,
    citecolor=blue,
    urlcolor=blue
}

\newcommand{\A}{\mathcal{A}}
\newcommand{\E}{\mathcal{E}}

\newcommand{\G}{\mathcal{G}}
\newcommand{\B}{\mathcal{B}}
\newcommand{\ba}{\begin{align}}
\newcommand{\ea}{\end{align}}

\def\3nab{\tilde{\nabla}}

\def\c{\mbox{curl}}
\def\be {\begin{equation}}
\def\ee {\end{equation}}
\def\ba {\begin{eqnarray}}
\def\ea {\end{eqnarray}}

\newcommand{\sfr}[2]
{{\textstyle\frac{#1}{#2}}}

\newcommand{\J}{{\mathcal J}}
\newcommand{\Q}{{\mathcal Q}}
\renewcommand{\H}{{\mathcal H}}
\newcommand{\barray}{\begin{array}}
\newcommand{\earray}{\end{array}}
\newcommand{\e}{e}
\newcommand{\N}{N}

\newcommand{\n}{{}^{(3)}\nabla}

\newcommand{\bea}{\begin{eqnarray}}
\newcommand{\eea}{\end{eqnarray}}
\newcommand{\f}[2]{\textstyle\frac{#1}{#2}}
\newcommand{\F}[2]{\frac{#1}{#2}}

\renewcommand{\S}{_{\mathsf{S}}}
\newcommand{\T}{_{\mathsf{T}}}
\usepackage[dvipsnames]{xcolor}

\begin{document}

\title[Integrated Cosmological memory]{Integrated cosmological memory: A dark-siren method to probe dark energy}

\author{Indranil Chakraborty}
\affiliation{Department of Physics, University of Virginia, Charlottesville, Virginia 22904-4714, USA}
\affiliation{Department of Physics, Indian Institute of Technology Bombay, Mumbai, Maharashtra 400076, India}
\altaffiliation{The first three authors contributed equally to this work.}
\email{indranil@virginia.edu}

\author{Sayantan Ghosh}
\affiliation{Department of Physics, Indian Institute of Technology Bombay, Mumbai, Maharashtra 400076, India}
\altaffiliation{These authors contributed equally to this work.}
\email{stanghosh@iitb.ac.in}

\author{Susmita Jana}
\affiliation{Asia Pacific Center for Theoretical Physics, Pohang 37673, Korea}
\affiliation{Department of Physics, Indian Institute of Technology Bombay, Mumbai, Maharashtra 400076, India}
\altaffiliation{These authors contributed equally to this work.}
\email{susmita.jana@apctp.org}

\author{Archana Pai}
\affiliation{Department of Physics, Indian Institute of Technology Bombay, Mumbai, Maharashtra 400076, India}
\email{archanap@iitb.ac.in}

\author{S. Shankaranarayanan}
\affiliation{Department of Physics, Indian Institute of Technology Bombay, Mumbai, Maharashtra 400076, India}
\email{shanki@iitb.ac.in}

\begin{abstract}
Gravitational-wave (GW) cosmology is currently bottlenecked by the scarcity of electromagnetic counterparts for bright sirens and the systematic uncertainties of galaxy catalogs for dark sirens. We propose a purely gravitational resolution using the Integrated Cosmological Memory (ICM)—the cumulative GW strain encoded in the spacetime geometry of an expanding Universe. While the GW transient emitted by the source provides the luminosity distance, the ICM accumulates a mathematically distinct integral of the cosmic expansion history. We demonstrate that extracting both observables from a single binary merger completely breaks the distance-redshift degeneracy within the gravitational sector. This establishes a novel, catalog-free dark siren framework for third-generation GW detector networks. Crucially, the resulting constraints on late-time dark energy are only weakly sensitive to the local expansion rate, providing a robust cosmological probe that can potentially mitigate the impact of the $H_0$ tension.
\end{abstract}

\keywords{gravitational waves --- cosmology --- dark energy --- dark siren --- gravitational wave memory} 

\acrodef{GWOSC}[GWOSC]{Gravitational Wave Open Science Center}
\acrodef{KDE}[KDE]{kernel density estimate}
\acrodefplural{KDE}[KDEs]{kernel density estimates}
\acrodef{BBH}[BBH]{binary black hole}
\acrodef{BNS}[BNS]{binary neutron star}
\acrodef{NSBH}[NSBH]{neutron star black hole}
\acrodef{BH}[BH]{black hole}
\acrodef{LVC}[LVC]{LIGO Scientific and Virgo Collaborations}
\acrodef{GW}[GW]{gravitational-wave}
\acrodefplural{GW}[GWs]{gravitational waves}
\acrodef{CBC}[CBC]{compact binary coalescence}
\acrodefplural{CBC}[CBCs]{compact binary coalescences}
\acrodef{CI}[CI]{confidence interval}
\acrodef{IMBH}[IMBH]{intermediate-mass black hole}
\acrodef{SMBH}[SMBH]{supermassive black hole}
\acrodefplural{SMBH}[SMBHs]{supermassive black holes}
\acrodefplural{IMBH}[IMBHs]{intermediate-mass black holes}
\acrodef{SNR}[SNR]{signal-to-noise ratio}
\acrodef{FAR}[FAR]{false alarm rate}
\acrodef{PSD}[PSD]{power spectral density}
\acrodefplural{PSD}[PSDs]{power spectral densities}
\acrodef{LVK}[LVK]{LIGO, Virgo and KAGRA}
\acrodef{GR}[GR]{General Relativity}
\acrodef{FF}[FF]{fitting factor}
\acrodef{O3}[O3]{third observing run} \acrodef{GWTC}[GWTC]{Gravitational Wave Transient Catalogue}  
\acrodefplural{GWTC}[GWTCs]{Gravitational Wave Transient Catalogues}
\acrodef{IFAR}[IFAR]{inverse false alarm rate}
\acrodefplural{IFAR}[IFARs]{inverse False Alarm Rates}
\acrodef{BHB}[BHB]{black hole binary}
\acrodefplural{BHB}[BHBs]{black hole binaries}
\acrodef{LHO}[LHO]{LIGO-Hanford}
\acrodef{LLO}[LLO]{LIGO-Livingston}
\acrodef{IGWN}[IGWN]{International Gravitational-Wave Observatory Network}
\acrodef{cWB}[cWB]{coherent WaveBurst}
\acrodef{PE}[PE]{parameter estimation}
\acrodef{CE}[CE]{Cosmic Explorer}
\acrodef{ET}[ET]{Einstein Telescope}

\section{Introduction} 

Gravitational wave (GW) astronomy has provided an unprecedented laboratory for testing the fundamental nature of gravity and the evolution of our Universe~\citep{Abbott:2016,Abbott:tests-of-GR}. Among its most transformative applications is the use of \acp{CBC}
as \emph{bright sirens}~\citep{Abbott:2017,Abbott:2017-H0,Abbott:MMA}. By providing a direct measurement of the luminosity distance ($d_L$) independent of the distance ladder, GWs offer a potentially clean resolution to the persistent tensions in precision cosmology, most notably the $H_0$ discrepancy between local and early-universe measurements~\citep{Abbott:2017-H0}.

However, the full potential of GW cosmology remains bottlenecked. \emph{Bright sirens}~\citep{Holz:2005,Abbott:2017-H0,Nissanke:2009} are hindered by the extreme scarcity of multimessenger events like GW170817~\citep{KAGRA:2021}, while \emph{dark sirens} \citep{Schutz:1986,Fishbach:2018,DES:2019} --- which rely on statistical associations with host-galaxy catalogs --- face significant systematic uncertainties due to catalog incompleteness and host-galaxy identification at high redshifts~\citep{dark-siren,Mastrogiovanni:2024}. 
Furthermore, the Hubble tension limits the reliability of cosmological probes that are strongly sensitive to variations in $H_0$~\citep{DiValentino:2021izs}. These challenges necessitate the search for a purely gravitational, catalog-independent probe that mitigates the systematic impact of $H_0$ while breaking current degeneracies to explore the expansion history of the Universe at cosmological scales.

In this article, we propose that the Integrated Cosmological Memory (ICM)~\citep{Chakraborty_IITB:2024,Chakraborty:2025qcu} --- embedded in the post-merger phase of the GW transient signal --- is the key to unlocking this probe. GW memory is fundamentally rooted in BMS symmetry and the soft graviton theorems of field theory~\citep{Strominger:2014, Flanagan:2015-Charges}. While traditionally studied in the idealized limit of asymptotically flat spacetimes~\citep{2023-Silvia.etal-PRD, 2024-Inchauspe-Silvia.etal-PRD, 2025-Silvia.etal-PRD,Tolish_Wald_cosmology:2016, Bieri_cosmology:2017, Chu:2016, Kehagias:2016}, recent theoretical advancements have established a rigorous framework for GW memory in Friedmann–Lema\^itre–Robertson–Walker (FLRW) geometries~\citep{Chakraborty_IITB:2024,Chakraborty:2025qcu}. 
{This formalism reveals that the memory strain contains a unique, cumulative contribution of the cosmological background. Conceptually analogous to the Integrated Sachs-Wolfe effect in the cosmic microwave background, the primary GW transient sources secondary GWs as it propagates across cosmological distances, accumulating a memory offset strictly dependent on the intervening expansion history:}
%
\begin{eqnarray}
\label{eq:GWmemory-FLRW}
\mathcal{N}_+ =
\frac{h_\oplus}{3 \, E^{2/3}(z_0)}
\int_0^{z_0} dz \, \frac{(1+z)}{E^{4/3}(z)}
\end{eqnarray}
where $z_0$ denotes the source redshift, $h_\oplus$ is the maximum intrinsic strain of the astrophysical transient GW signal emitted by the source measured at the detector and $E(z) \equiv H(z)/H_0$ (dimensionless Hubble expansion rate) characterizes the late-time expansion model. We refer to this integral as the ICM.

\begin{figure*} 
    \centering
    \includegraphics[width=\linewidth]{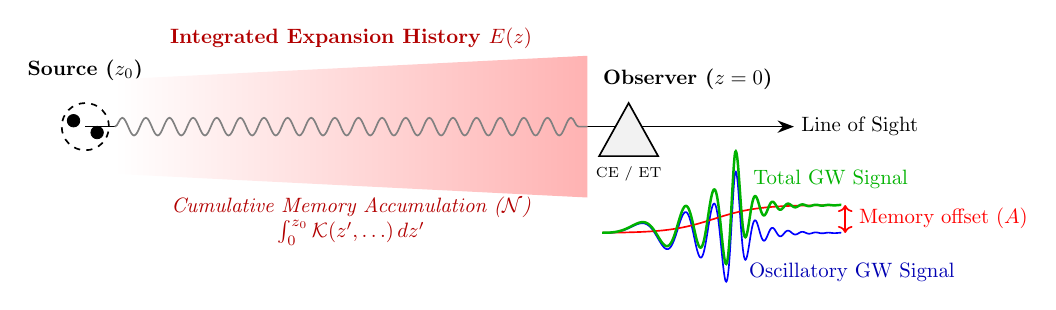}
    \caption{
    {\bf Schematic of GW propagation and ICM accumulation.
    }Signals propagate from source ($z_0$) to observer ($z=0$), where the redshifted intrinsic GW signal from the source, depicted as an oscillatory wave (blue), encodes the luminosity distance $d_L$. The memory strain (red) accumulates through the merger to a saturated offset, resulting in the total observed signal (green). This ICM accumulation depends on the integrated expansion history $E(z)$ via the kernel $\mathcal{K}$ (shaded), providing a purely gravitational probe of background cosmology.}
    \label{fig:ICM}
\end{figure*}

Traditional standard sirens are limited as the GW transient emitted by the source provides only $d_L(z)$, leaving the source redshift $z$ perfectly degenerate. The methodological breakthrough of this work lies in recognizing that the ICM (Eq.~\ref{eq:GWmemory-FLRW}) accumulates a mathematically distinct integral of the expansion history. By extracting both $d_L$ and the ICM offset for the same GW event, we enable a novel dual-observable approach. Here, for the first time, we show that it is \emph{possible to break} the distance-redshift degeneracy \emph{weakly dependent} on the local expansion rate, purely from the morphology of a single GW event. As shown in Fig.~\ref{fig:ICM}, this transforms the memory offset into an \emph{independent cosmological observable}.

The significance of ICM lies in its distinct functional dependence on cosmological parameters, which differs fundamentally from that of the luminosity distance $d_L (z)$. While the transient GW signal amplitude scales inversely with the luminosity distance $d_L(z)$ (including $(1+z)$ redshift dependence), the ICM introduces an additional, non-trivial integrated kernel (see Fig.~\eqref{fig:ICM} and Eq. \eqref{eq:GWmemory-FLRW}) that is sensitive to the expansion history $E(z)$ in a mathematically distinct way. Capitalizing on this, we develop independent probes to estimate both the GW source parameters and the ICM memory components from a single event. The synergy between these two independent gravitational observables has the potential to break the parameter degeneracies among the $d_L$ and $E(z)$ that typically limit standard sirens~\citep{Xu:2024}. 

In this work, we provide a proof-of-principle demonstration that a network of next-generation (nG) ground-based detectors --- specifically the Einstein Telescope (ET) and Cosmic Explorer (CE) --- is sufficiently sensitive to place meaningful, purely gravitational constraints on the ICM and the late-time expansion history without any input from EM observations. Using an injection campaign containing full astrophysical GW waveforms (compact binary transients plus the ICM) into simulated nG detector noise, we isolate the cosmological contribution to the memory strain:
\begin{equation}
\label{eq:hmem}
h_{\rm mem}(t,z_0) = \mathcal{N}(z_0) \sigma(t);~
\sigma(t) = [1+e^{-(t-t_0)/\tau}]^{-1} \, ,
\end{equation}
where $\mathcal{N}(z_0)$ depends on the background cosmology, $t_0$ is the merger time, and $\tau$ represents the memory rise-time (set to $0.01$s)\footnote{{The rise time scales roughly as the dynamical time for energy radiation during merger $t_{dyn}\sim GM/c^3$ where $M$ is the total mass~\citep{Thornenature1987} The total mass of the source considered in this work $\sim 300 M_\odot$. This gives roughly $t_{dyn} \sim 0.01s$.}}. Since, GW detectors lack sensitivity to a DC offset, the ICM signal observed in the detector's noise will be a high-passed version of the step-like ICM signal \citep{Lasky:2016}, appearing as a short-duration burst. Thus, a dedicated transient search can indirectly estimate the memory ratio, differentiating cosmological models.

\begin{figure*}[!htb]
    \centering
    \includegraphics[width=0.6\linewidth]{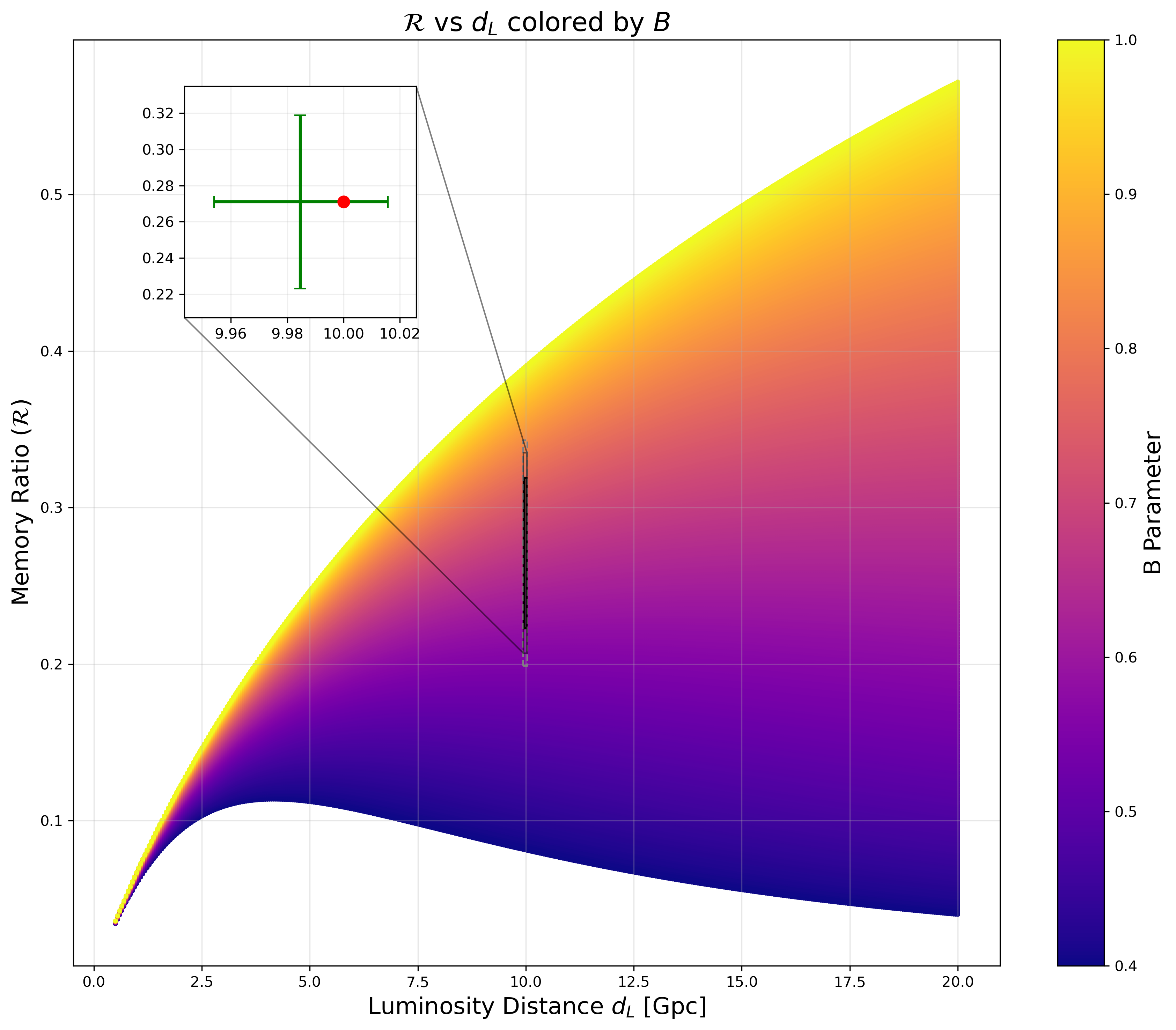}
    \caption{{\bf The ICM memory ratio $\mathcal{R}$ as a function of $d_L$ (for $H_0=70 \,\, \text{km} \,\,\text{s}^{-1}\text{Mpc
    }^{-1}$) and the expansion parameter $B$ (color scale).} Inset:  The 90\% credible intervals of $d_L$ and $\mathcal{R}$ are denoted by the solid green lines. The simulated injection at $d_L = 10\,\mathrm{Gpc}$ assuming $\Lambda$CDM ($B = 2/3$) is shown by the red dot. }
    \label{fig:R-dL-B}
\end{figure*}

\section{Methods}\label{sec2}

\noindent In the detector bandwidth, the GW strain is the sum of the CBC signal $h_{cbc}(t)$, the detector noise $n(t)$ and the high-passed \footnote{We use the zero-phase digital filter -- a 8th-order Butterworth filter with a 5 Hz cutoff implemented using the \texttt{pycbc.filter.resample.highpass} module.} ICM signal  $h^{\mathrm{HP}}_{\mathrm{mem}}(t)$~\citep{Favata:2010}:
\begin{equation}{\label{eqn:data-noise-signal}}d (t) = n(t) + h_{\mathrm{cbc}}(t) + h^{\mathrm{HP}}_{\mathrm{mem}}(t) \,.
\end{equation}

\subsection*{Parameter recovery for CBC source}

We use standard Bayesian Parameter Estimation (PE)~\citep{veitch, thrane} to obtain the CBC parameters using the Bilby package \citep{Ashton2019Bilby}. The posterior distribution of the CBC source parameters $\vec{\theta}$ is $p(\vec{\theta}|d) \propto \mathcal{L}(d|\vec{\theta}) \pi (\vec{\theta})$, where $\mathcal{L}$ is the likelihood function and $\pi$ is the prior. This procedure yields a set of posterior samples in a multi-dimensional astrophysical parameter space representing our best reconstruction of the source parameters (including $d_L$) for a given waveform model, $h_{+/\times}(t, \vec{\theta})$ \citep{veitch, thrane}. Current GW memory literature~\citep{Lasky:2016,Hubner:2020,Khera:2020,Boersma:2020} often use waveform models with astrophysical memory components, thus enabling estimation of all signal parameters (memory included) at a single go. Here, we use ICM rather than astrophysical memory and thus, propose to estimate the source and ICM parameters using two independent approaches. 
Further, at present there are no waveform models which include both the \ac{CBC} signal and the ICM signal.

As we show later, in the  nG detector network,  the CBC source parameters are not biased in presence of ICM. This enables a fairly accurate reconstruction of the CBC signal and its subsequent subtraction from the data to give us the residual $d_{\mathrm{sub}}(t) = n(t) + h^{\mathrm{HP}}_{\mathrm{mem}}(t) + \delta h_{cbc}(t) $\footnote{Since the highpassed memory transient has a fundamentally different time-frequency morphology than the chirping leakage~\citep{Davis:2018yrz}, the cross-term $(\delta h_{cbc}|\hat{h})$ is highly suppressed, allowing us to treat the leakage as sub-dominant to the Gaussian noise variance.}, where $\delta h_{cbc}(t)$ is the difference between the true signal and the maximum likelihood waveform, which we assume to be small for accurate waveform reconstruction. Next, we use the matched-filtering technique to obtain the ICM amplitude signal~\citep{Favata:2010,Talbot:2018}.  We express the high-passed memory transient as: $h^{\mathrm{HP}}_{\mathrm{mem}}(t) = a \hat{h}$, where $\hat{h}$ is normalized ($(\hat{h} | \hat{h}) = 1$) and $a(A)$ is the optimal SNR and is a function of the memory offset; $A$. Matched-filtering the residual data against the memory template, we get: $(d|\hat{h}) = (n|\hat{h}) + a$ (dropping subscripts and superscript for brevity). For stationary Gaussian noise, $(d|\hat{h}) \sim \mathcal{N}(a,1)$ and thus, we obtain the point estimate $\hat{A}$ and the associated errors \footnote{The variables with hat are the estimates of the unhatted parameters.}. For more details, see the Appendix C. 
 
In PE, we use the network of nG GW detector networks and the waveform model \texttt{IMRPhenomXAS} that contains only the intrinsic signal emitted by the CBC source, without any ICM component.
By estimating the CBC amplitude $h_\oplus$ (from PE) and the offset ${\hat A}$ from matched-filtering, we determine the total memory ratio $\mathcal{\hat{R}} = {\hat{A}}/h_\oplus$.
In Einstein's GR, the observed GW strain is a superposition of the plus ($+$) and cross ($\times$) polarization components, projected onto the detector's antenna response functions $F_+$ and $F_\times$. 
The time-dependent memory signal is $h_{\mathrm{mem}}(t) = F_+ h^+_{\mathrm{mem}}(t) + F_\times h^\times_{\mathrm{mem}}(t)$. 
If we consider $A_+$ and $A_\times$ to be memory off-sets in two polarizations, for a given detector pair, we can estimate them as follows. Let us consider two detectors (1 and 2) with their respective estimates of the memory offset as
\begin{equation}
{\hat{A}}_1 = F_{+1} A_+ + F_{\times1} A_\times\, ,~
{\hat{A}}_2 = F_{+2} A_+ + F_{\times2} A_\times \, .
\end{equation}
Inverting the above equations \footnote{This can be done provided the antenna pattern matrix is non-singular (which happens for misaligned detectors). For any non-aligned detector combination, barring a few directions where the determinant may become small, this inversion will provide a unique solution.}, we get:
\begin{equation}{\label{eqn:F_matrix_inversion}}    
\begin{bmatrix}
{\hat A}_+ \\
{\hat A}_\times
\end{bmatrix}
=
\begin{bmatrix}
F_{+1} & F_{\times 1} \\
F_{+2} & F_{\times 2}
\end{bmatrix} ^ {-1}
\begin{bmatrix}
{\hat A}_1 \\
{\hat A}_2
\end{bmatrix}
\end{equation}
\begin{figure*}[t]
    \centering
    \begin{minipage}{0.32\textwidth}
        \centering
        \includegraphics[width=\linewidth]{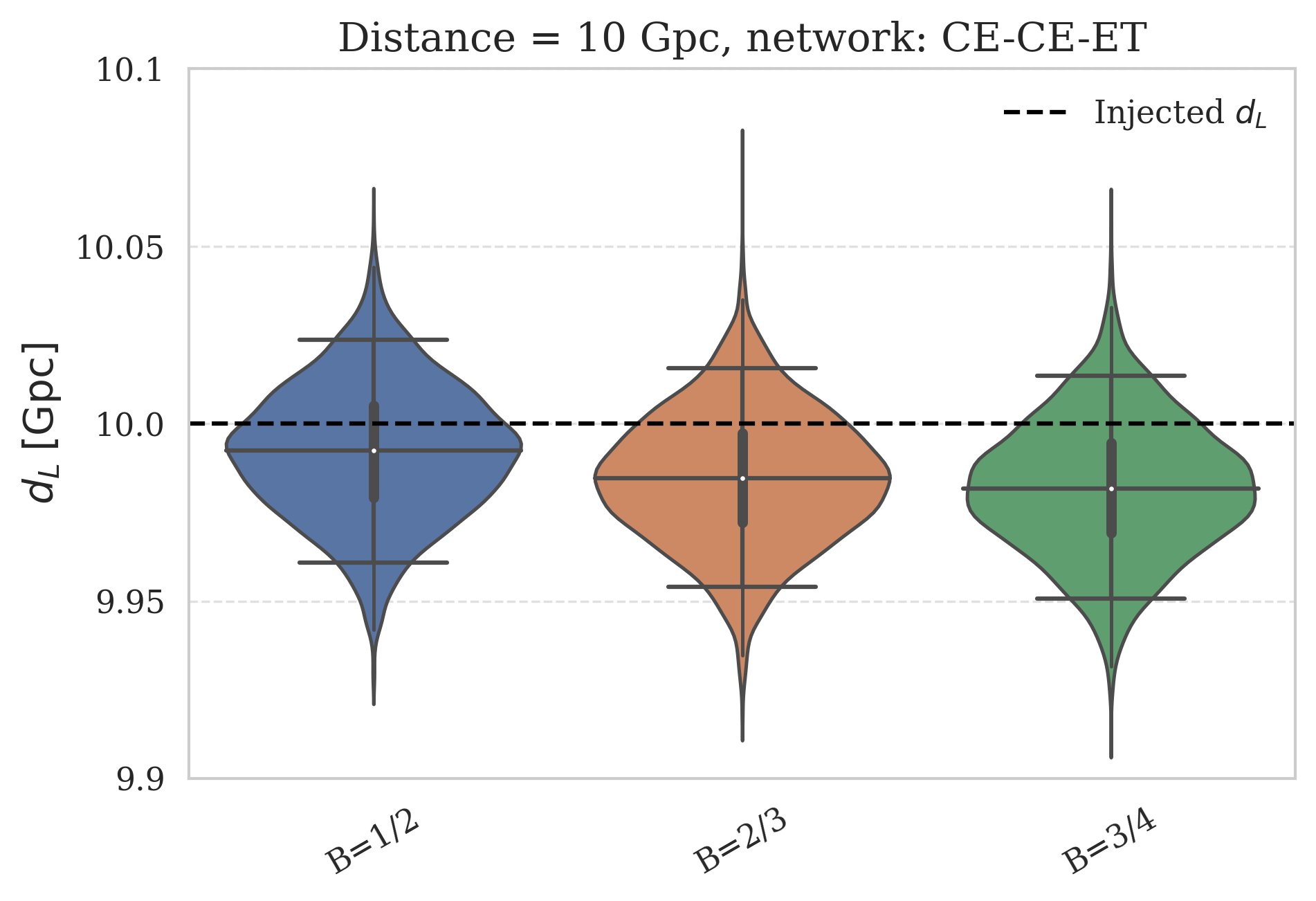}
    \end{minipage}\hfill
    \begin{minipage}{0.32\textwidth}
        \centering
        \includegraphics[width=\linewidth]{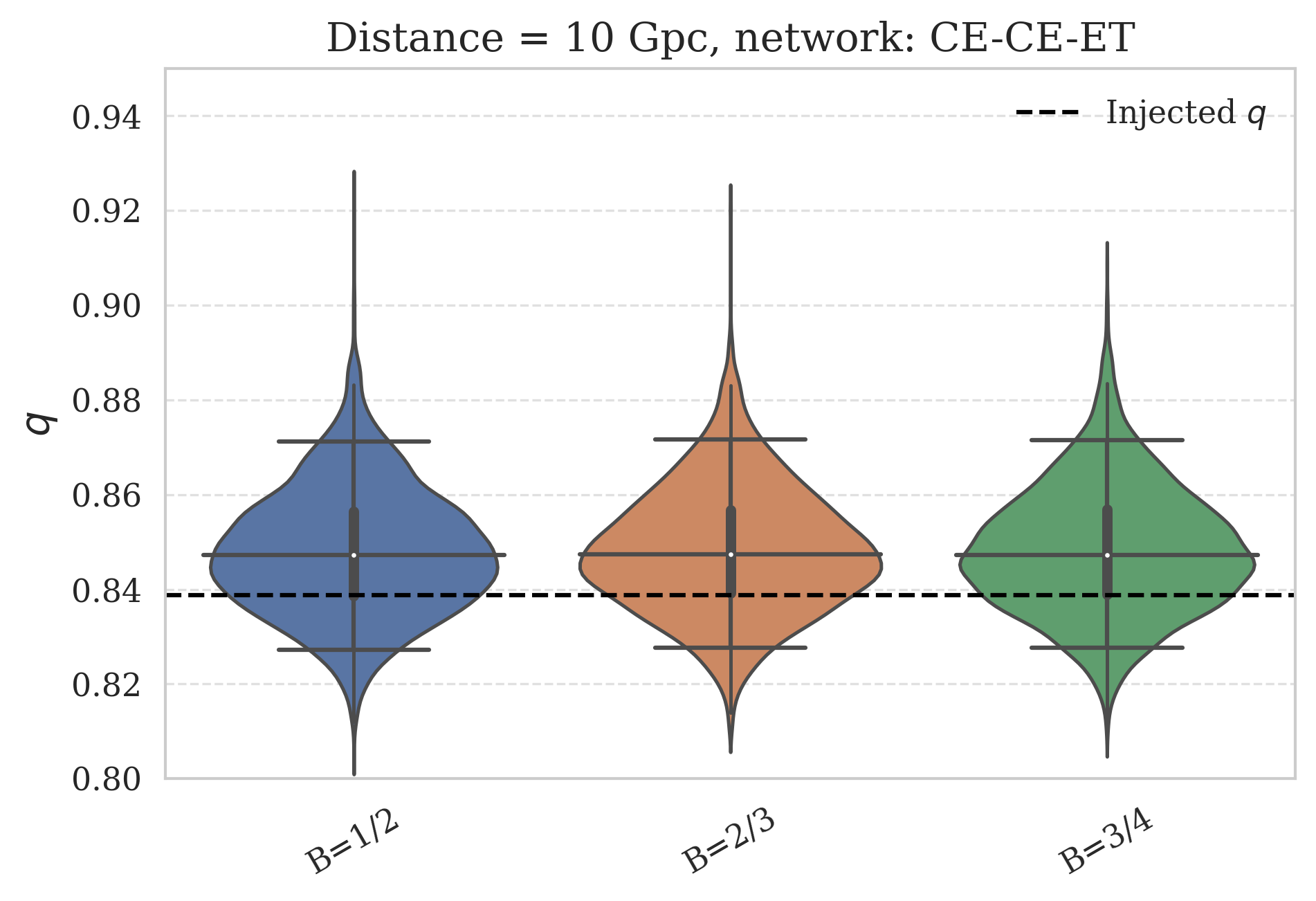}
    \end{minipage}\hfill
    \begin{minipage}{0.32\textwidth}
        \centering
        \includegraphics[width=\linewidth]{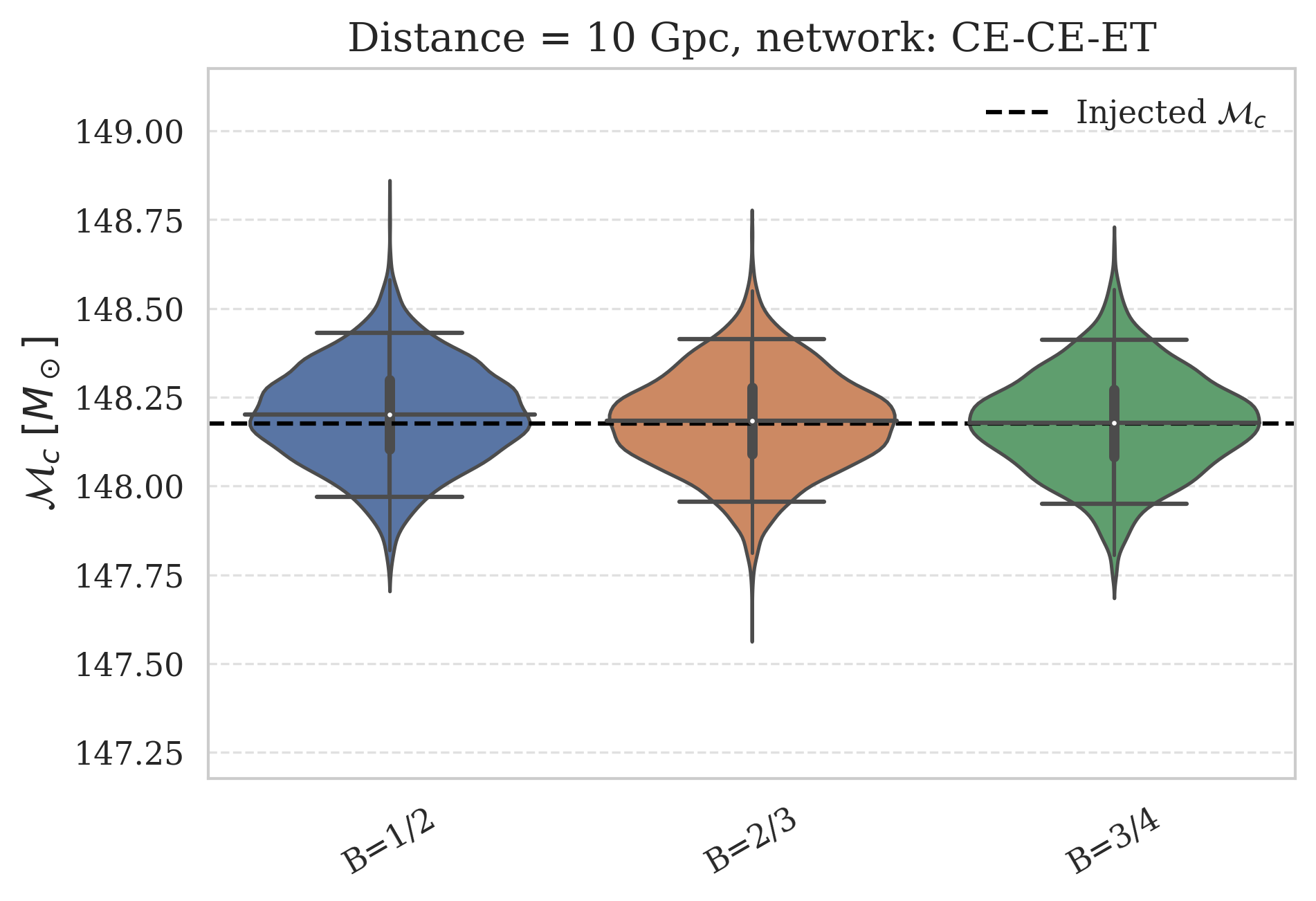}
    \end{minipage}
    \caption{{\bf Parameter recovery for the CE-CE-ET network at $10$ Gpc (for $H_0=70 \,\, \text{km} \,\,\text{s}^{-1}\text{Mpc
    }^{-1}$).} Posteriors for $d_L$, mass ratio $q$, and chirp mass $\mathcal{M}_c$. In each panel, the violin plots represent the three expansion histories ($B = 1/2, 2/3, 3/4$). In all cases, the true parameters are recovered within $90\%$ credible intervals, demonstrating that the memory signal does not bias source parameters. For other configurations and distances, see the appendix B.}
    \label{fig:3x3_fullpage_posteriors_cbc_parameters}
\end{figure*}

From the PE posterior samples,
we construct resulting CBC waveforms - one for each sample point. We further obtain the $+$ and $\times$ polarization and estimate their respective peak amplitudes, $({\hat h}_{+,\text{peak}}^{(i)}, {\hat h}_{\times,\text{peak}}^{(i)})$. Thus, we reconstruct the polarization-dependent memory ratios:
\begin{align}\mathcal{{\hat R}}_+^{(i)} = {\hat A}_+ / {\hat h}_{+,\text{peak}}^{(i)} , \quad \mathcal{{\hat R}}_\times^{(i)} = {\hat A}_\times / {\hat h}_{\times,\text{peak}}^{(i)} \, .
\end{align}
To mitigate the impact of residual noise and provide a robust metric for cosmological inference, for each sample, we estimate $\mathcal{{\hat R}} = (\mathcal{{\hat R}}_+ + \mathcal{{\hat R}}_\times)/2$, thus translating the $\mathbf{\theta}$ posterior to an inferred $\mathcal{R}$ distribution \footnote{We can loosely call this the $\mathcal{R}$ ``posterior'', but it is not truly a posterior, as it is obtained by a combination of Bayesian and frequentist point estimates} . Finally, to map the recovered $90\%$ credible intervals of $\mathcal{R}$ and $d_L$ onto the $B$-parameterized theoretical surface as illustrated in the $\mathcal{R}-d_L$ plane in Fig.~\eqref{fig:R-dL-B}, we use one specific value of $H_0 = 70$~{km/s/Mpc} and we show that the results are weakly dependent of the choice of $H_0$. 
Due to the smooth variation of $B$ across this parameter space, a sufficiently high SNR in future detectors allows this region to shrink, thereby providing purely gravitational constraints on
$B$. We emphasize that this procedure serves to constrain $B$ rather than estimating it as a model parameter in the strict Bayesian sense. 

\section{Results}\label{sec4}

\noindent As a demonstration, we use a nG detector network comprising two CE observatories (located at LIGO Hanford and Livingston sites), and an ET (at the Virgo site). A three-detector network is essential for accurate sky localisation and identification of the two GW polarisations. For memory estimation and probing cosmology, accurate source localization is critical. Large uncertainties in the sky location propagate into the $d_L$ estimates, thus broadening the constrained region used for inferring ${\hat B}$. For the three-detector network, we compute ${\cal {\hat R}}$ for that pair of detectors which gives highest memory SNRs, for instance, for the CE-CE-ET network, we consider the two CE detectors for memory estimation \footnote{At the distances considered in this work, the memory SNRs for the ET detector are not high enough to be used for memory ratio estimation, although they play a significant role in sky localization.}. 

We inject a \ac{BBH} binary signal into stationary Gaussian noise colored by the analytical power spectral densities (PSDs) of the respective detectors. 
We consider a non-spinning, quasi-circular, face-on \ac{BBH} 
binary with detector-frame component masses $m_1 = 186 M_{\odot}$ and $m_2 = 156 M_{\odot}$, at a sky location ($\mathrm{RA} = 0.022, \mathrm{DEC} = -0.90$), at which the matrix in Eqn. \ref{eqn:F_matrix_inversion} is invertible and has a high antenna pattern response (see the appendix B for more details).
{For this configuration, the intrinsic (Christodoulou) nonlinear memory identically vanishes ensuring that the injected memory is entirely zero at the source. Consequently, any memory extracted at the detector is guaranteed to be the pure ICM generated during propagation, allowing for a clean, uncontaminated test of the background cosmology.}

We use \texttt{IMRPhenomXAS} approximant \citep{IMRPhenomXAS} for GW signal from the CBC source, while the memory component is the highpassed sigmoid $\sigma(t)$ function as explained earlier. As seen in Fig.(\ref{fig:R-dL-B}), $\mathcal{R}$ is low and degenerate across cosmologies at small distances. Conversely, at $d_L \gtrsim 10$~Gpc, $\mathcal{R}$ is both enhanced and cosmologically distinct.  Thus, we choose signals for performing
injections at three  cosmological distances ($5, 8, 10$ Gpc) (corresponding to $z \approx (0.8,1.2, 1.4)$ for $\Lambda$CDM). Additionally, we consider three distinct expansion histories parameterized by $B = \{1/2, 2/3, 3/4\}$.
These three $B$ values correspond to different late-time evolution --- $B = 1/2, 2/3$ and $3/4$ correspond to quintessence~\citep{Poulin:2018cxd}, $\Lambda$CDM and interacting dark energy~\citep{DiValentino:2021izs,Bansal:2025usn}, respectively.
Note that for convenience, we consider sources with fixed detector-frame masses which implies that their source-frame masses will differ from one case to another depending upon the luminosity distance and the cosmology.  

Fig. \ref{fig:3x3_fullpage_posteriors_cbc_parameters} shows the representative posterior distributions at $d_L=10$ Gpc. 
We find that the presence of the sub-dominant memory signal in the data \emph{does not bias} the recovery of the 
astrophysical parameters; the injected chirp mass $\mathcal{M}_c$, mass ratio $q$, and $d_L$ are 
consistently recovered within $90\%$ credible intervals across all distances and network configurations (see the appendix  B for the complete set of results). Although not shown here, the sky localization of the source is typically less than 1 sq. degree.

\begin{figure*}
    \centering
    \includegraphics[width=0.88\linewidth]{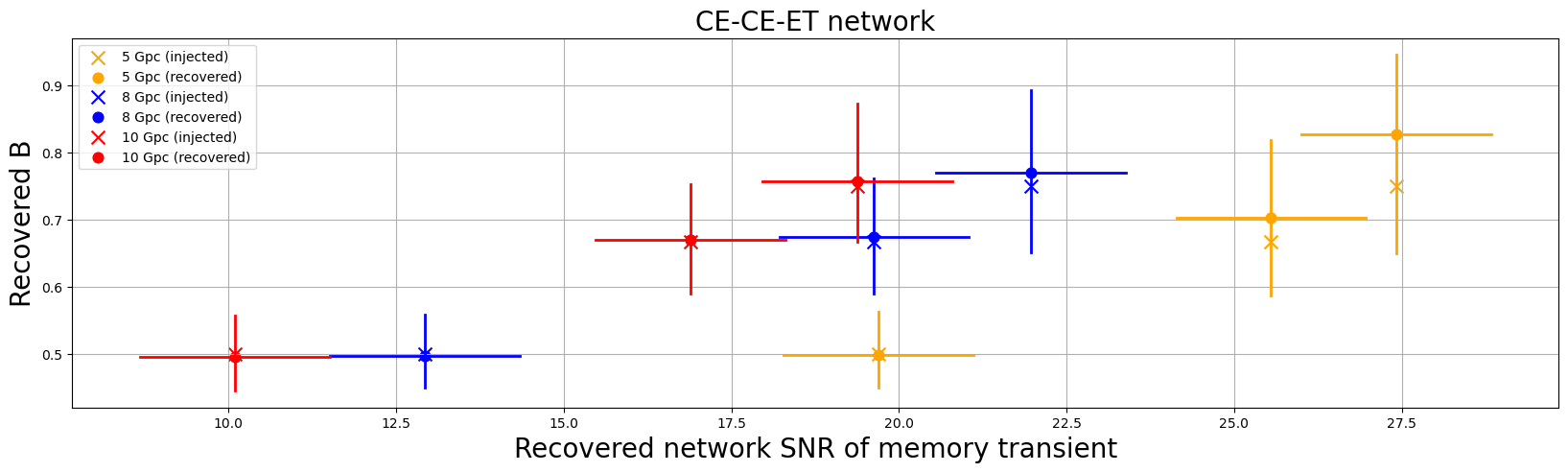}
    \caption{{\bf The x-axis shows the recovered network SNR (with uncertainties arising because of the stochastic nature of detector noise) of the memory signal for each distance and cosmology.} The y-axis shows the recovered $B$ constraints, along with their percentage deviations from the injected $B$ ($0.5$, $0.66$ and $0.75$)  values.}
    \label{fig:B_vs_SNR}
\end{figure*}

We combine $d_L$ posteriors \footnote{Please refer to the appendix B for the posterior distributions corresponding to signals injected at distances of 5, 8 and 10 Gpc. A comparison of the PE results at these distances is also provided. Nevertheless, in all cases, the parameter $B$ is recovered with reasonable precision} with the $\mathcal{\hat{R}}$ constraints, and obtain constrains on $B$ as summarized in Fig. \ref{fig:B_vs_SNR}. The intersecting sticks shown in the same colour correspond to a fixed distance and differ in the cosmology. We compare estimates from $\Lambda$CDM cosmology ($B = \frac{2}{3}$) with other cosmologies $B=1/2, 3/4$. We present results with a fiducial value of $H_0 = 70$ km/s/Mpc \citep{Abbott:2017-H0}. Note the analysis with two other values of $H_0=\{67,73\}$ km/s/Mpc \citep{planck-H0,shoes-H0}, shows that the constraints on  $B$ values differ by $\leq 4\%$. \footnote{Please see the appendix B for details}. {This indicates that the recovery of the parameter $B$ is largely insensitive to current uncertainties in $H_0$. The variations induced by changing $H_0$ are significantly smaller than the statistical uncertainties of the measurement.} In general, considering all the cases, the percentage accuracy of ${\hat B}$ estimate varies between $12-16\%$. Interestingly, for each distance, we observe that the recovered $B$ constraints for $\Lambda$CDM cosmology ($B=2/3$) are well separated from the quintessence model ($B=1/2$). However, there is a fair amount of degeneracy in the recovered distributions of $B=2/3$ and $B=3/4$, which clearly increases for smaller distances. Although SNR increases with decreasing distance, the differences between memory ratios corresponding to the three cosmologies diminish (see Fig. \ref{fig:R-dL-B}), thus reducing the distinguishability between the models at a fixed distance. We emphasize that these baseline constraints are derived from a single GW event; since nG detectors are projected to observe multiple mergers per day, a collective stacking analysis is expected reduce these observational error bars, a prospect currently under active investigation.

\section{Discussion}

\noindent By extracting $B$ from ICM, our methodology provides a purely gravitational, high-redshift lever to distinguish between these competing physical paradigms without relying on electromagnetic calibrations or low-redshift Taylor expansions (e.g., the CPL parametrization). However, to contextualize the ICM amid current observational tensions, most notably the preference for dynamic dark energy in DESI data~\citep{Abdalla:2022yfr,CosmoVerseNetwork:2025alb,Scott:2018adl,Peebles:2025lmy,DESI-DR2:2025}, we map our phenomenological parameter $B$ to the equation of state (EoS) $w(z)$. For a generic scalar-field model where the dimensionless Hubble parameter evolves as $E_{\rm DE}^{2}(z) = \Omega_{m0}(1+z)^{3} + (1-\Omega_{m0}) \exp\left[ 3\int_{0}^{z} \frac{1+w(z')}{1+z'} dz' \right]$, matching our parameterization $E(z)= \sqrt{\omega_{m0}(1+z)^{2/B}+(1-\Omega_{m0})}$ yields the effective EoS~\citep{PURBA-ANJAN:2025}:
\begin{equation} \label{eq:w-model}
w(z; B) = \frac{ \frac{2 \Omega_{m0}}{3B} (1+z)^{2/B} - E^{2}(z) }{E^{2}(z) - \Omega_{m0}(1+z)^{3}} \, .
\end{equation}
Performing a low-redshift Taylor expansion of Eq.~(\ref{eq:w-model}) to the standard Chevallier-Polarski-Linder (CPL) form $w(z) \approx w_0 + w_a z/(1+z)$ \citep{Chevallier:2000-CPL1,Linder:2002-CPL2}, we extract $(w_0, w_a)$ directly from $B$. For injected $\Lambda$CDM ($B=2/3$), the expected limit is $(-1, 0)$. Using our recovered $B$ distributions, we infer median CPL parameters $(w_0, w_a)$ of $(-1.022, -0.131)$ at $d_L = 5\,$Gpc, 
$(-1.005, -0.032)$ at $d_L = 8\,$Gpc, 
$(-1.002, -0.013)$ at $d_L = 10\,$Gpc.

{We note that the apparent systematic offset in the median of $w_a$ from $0$ is not a bias in the GW parameter estimation, but rather a known statistical artifact of projecting a fundamentally non-linear parameterization ($B$) onto a linearized, empirical Taylor expansion (CPL). Because the transformation from $B$ to $w_a$ is highly non-linear, standard Gaussian noise fluctuations in the $B$ posterior preferentially stretch the tails of the transformed $w_a$ distribution, shifting the median. Furthermore, because our injected GW sources extend to $z \approx 1.4$, we are probing a cosmological depth where the low-redshift CPL approximation natively loses its validity.} Consequently, this demonstrates that the deep reach of the ICM demands higher-order, non-parametric reconstructions, or native non-linear parameterizations like $B$, analogous to the requirements for high-redshift JWST observations~\citep{2025-Finkelstein-ApJL,Perlmutter2019Key,Ferraro2019Inflation}.
Consequently, the ICM serves as a powerful dark-siren alternative where EM standard candles become scarce or systematic-dominated.

Furthermore, the ICM provides a novel pathway to distinguish dynamic dark energy from modified gravity (MG)~\citep{Shankaranarayanan:2022wbx,Mandal:2025xuc,Slosar2019Dark,Bansal:2025usn}. MG theories typically alter GW propagation via friction or modified dispersion~\citep{Mandal:2025xuc,Mandal:2025xza}. Because the ICM is an integrated history of the wave's path, it naturally accumulates these deviations, imprinting distinct signatures, like enhanced polarization asymmetry in Chern-Simons gravity~\citep{Chakraborty:2025qcu}, onto the $\mathcal{R}-d_L$ relation. 

In this article, we have provided a proof-of-principle demonstration for a new class of purely gravitational cosmological probes. While standard EM-based standard candles currently provide tighter bounds on the dark energy equation of state, they are rapidly approaching a systematic floor. Our method provides a highly complementary, systematic-independent alternative. While a single golden event in the CE-ET network yields a $12-16\%$ constraint on the expansion parameter $B$, the true potential of this methodology can be unlocked with multiple events in the nG era. As nG networks detect thousands of high-SNR mergers, the joint hierarchical stacking of GW signals from a multitude of astrophysical sources will provide a transformative, galaxy catalog-free gravitational probe of fundamental physics on cosmic scales.

Lastly, a critical data-analysis consideration is ensuring the morphological separation of the post-merger memory transient from the quasi-normal mode (QNM) ringdown of the remnant black hole. Filtering the simulated band-limited memory signal against a bank of arbitrary damped sinusoids yields a highly statistically significant match ($0.99$). However, mapping the empirical frequency and damping time of this best-fit sinusoid to standard black hole remnant relations~~\citep{Berti:2005ys,berti-fits} yields an unphysically low quality factor ($Q \approx 0.47$). Standard BH ringdowns strictly require $Q \gtrsim 2$. This physical mismatch means that a matched-filter search designed to extract the memory transient will not accidentally lock onto the regular CBC ringdown. Consequently, the risk of residual ringdown leakage corrupting the memory extraction is naturally mitigated. 

\begin{acknowledgments}
The authors are grateful to V. Bhalerao, P. G. Christopher, J. Fernandes, S. Jana, O.~Hannuksela, K. Hari, and R. Kashyap for their valuable discussions and feedback on the earlier draft. 
IC is supported by the Fulbright-Nehru Postdoctoral Research Fellowship (Award No.3141/FNPDR/2025) from the United States-India Educational Foundation. SG acknowledges fellowship support provided by MHRD, Government of India. The work of SJ is supported by the Young Scientist Training  program at the Asia Pacific Center for Theoretical Physics.
AP acknowledges the support from SPARC MoE grant SPARC/2019-2020/P2926/SL, Government of India and ANRF MATRICS grant ANRF/ARGM/2025/001361/TS.
The work is supported by ANRF Advanced Research grant (ANRF/ARG/2025/001514/PS). 
\end{acknowledgments}

\begin{contribution}

IC played a leading role in formulating the ICM-based dark energy framework, wrote significant parts of the manuscript, developed the cosmological extraction pipeline, and contributed to the parameter estimation.

SG played a leading role in developing the method for extracting memory ratio from the data and combining the memory ratio estimates with source parameter estimates to constrain cosmological model parameters. SG also carried out most of the injection campaign including the parameter estimation runs and wrote significant parts of the manuscript. 

SJ played a leading role in formulating the ICM-based dark energy framework, wrote significant parts of the manuscript, developed the cosmological extraction pipeline, and contributed to the parameter estimation.

AP and SS supervised the whole project, and wrote significant portions of the manuscript.


\end{contribution}

\appendix

\section{Physical Interpretation of the Expansion Parameter $B$}

In the main Letter, we adopted a phenomenological parametrization for the dimensionless Hubble parameter $E(z) \equiv H(z)/H_0$, introduced by Sen and Sethi~\citep{Sen_PLB:2001}:
\begin{equation} \label{eq:Ez_supp}
    E^2(z) = \Omega_{m0} (1+z)^{2/B} + (1-\Omega_{m0}) \, .
\end{equation}
Rather than assuming a specific dark energy equation of state \textit{a priori}, this parametrization modifies the scaling of the high-redshift component. 
Depending on the value of $B$, the cosmological history maps precisely onto several well-motivated theoretical paradigms:
\begin{description}
    \item[$\Lambda$CDM ($B = 2/3$)] For $B = 2/3$, the exponent reduces to $3$. Eq.~\ref{eq:Ez_supp} directly recovers the standard spatially flat $\Lambda$CDM expansion history:
\begin{equation}
    E^2(z) = \Omega_{m0} (1+z)^3 + (1-\Omega_{m0}) \, ,
\end{equation}
where the first term represents pressureless dust (dark matter and baryons) diluting with the expanding volume, and the second term is the constant cosmological constant ($\Lambda$).

\item[Quintessence ($B < 2/3$)] When $B < 2/3$, the high-redshift component of the universe dilutes faster than standard matter. For instance, for $B = 1/2$ (which we study in this work), the exponent becomes $4$, yielding an expansion history driven by an effective radiation-like term at early times:
\begin{equation}
    E^2(z) = \Omega_{m0} (1+z)^4 + (1-\Omega_{m0}) \, .
\end{equation}
When mapped to an effective dark energy equation of state $w(z)$ (as derived in the main text), this scenario corresponds to $w > -1$. Physically, this represents a {\it Quintessence} model, where dark energy is sourced by a slowly rolling scalar field. Alternatively, at higher redshifts, it mimics {\it Early Dark Energy} models~\citep{Poulin:2018cxd}, where the dark sector injects energy early in cosmic history before diluting rapidly (akin to radiation) to leave the late-time universe matter-dominated. 

\item[Interacting Dark Sectors ($B > 2/3$)]
Conversely, for $B > 2/3$, the high-redshift component dilutes slower than standard non-relativistic matter. For $B = 3/4$ (studied in this work), the exponent becomes $8/3 \approx 2.66$:
\begin{equation}
    E^2(z) = \Omega_{m0} (1+z)^{8/3} + (1-\Omega_{m0}) \, .
\end{equation}
Mapping this to the effective dark energy framework yields $w < -1$, characteristic of {\it Phantom Dark Energy} models that violate the null energy condition \citep{Ludwick:2017}.

Equally compelling is the interpretation of $B > 2/3$ through the lens of an {\it Interacting Dark Sector}. If dark matter and dark energy are non-minimally coupled such that dark matter decays into dark energy as the universe expands, the matter density will not fall off as steeply as $(1+z)^3$. The parameter $B=3/4$ provides a phenomenological fit to such interacting fluid dynamics, which have recently gained significant traction as a potential resolution to the $S_8$ and $H_0$ tensions \citep{DiValentino:2021izs,Bansal:2025usn}.
\end{description}

\begin{figure*}[htb!]
    \centering

    \begin{minipage}{0.3\textwidth}
        \centering
        \includegraphics[width=\textwidth]{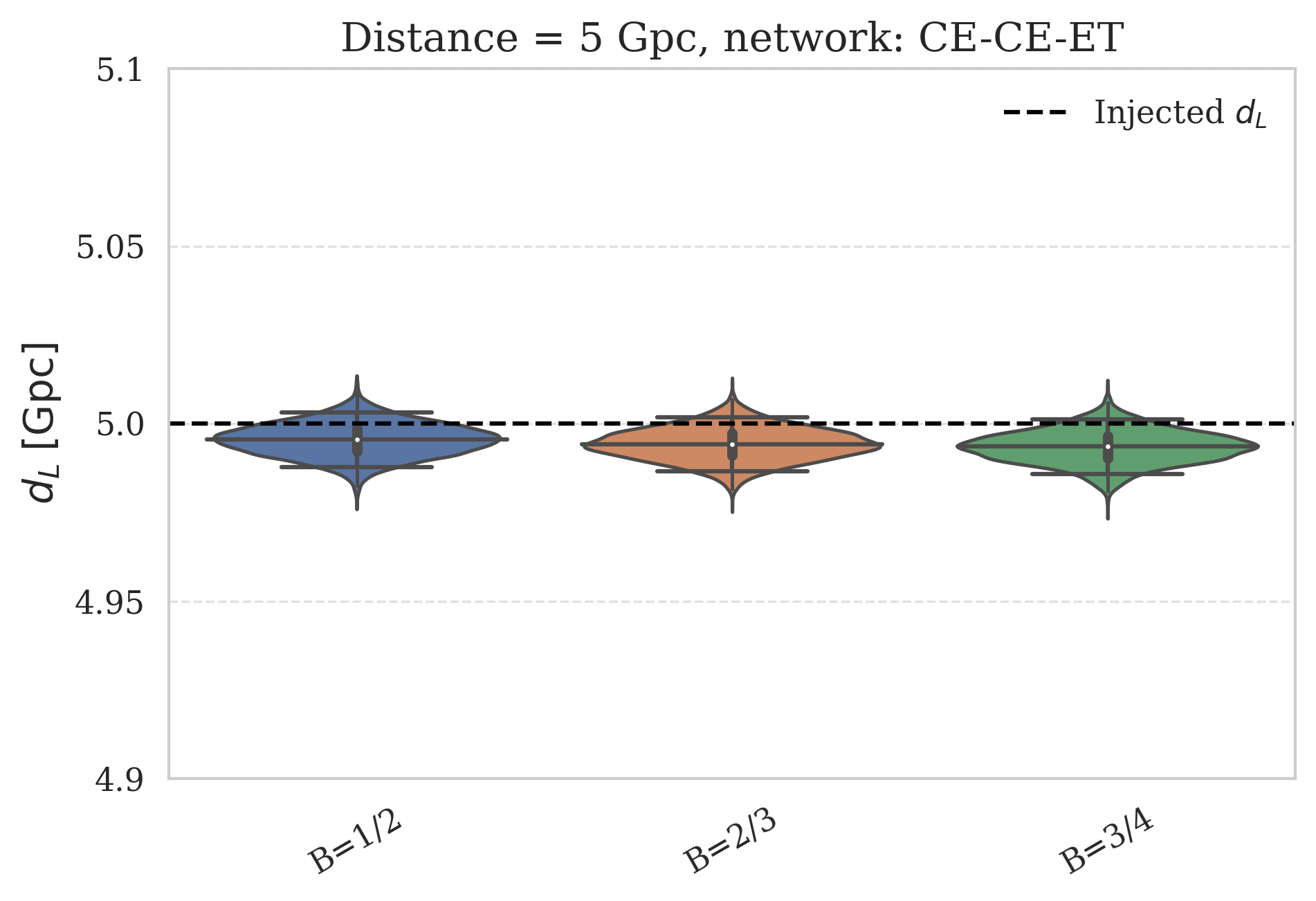}
    \end{minipage}\hfill
    \begin{minipage}{0.3\textwidth}
        \centering
        \includegraphics[width=\textwidth]{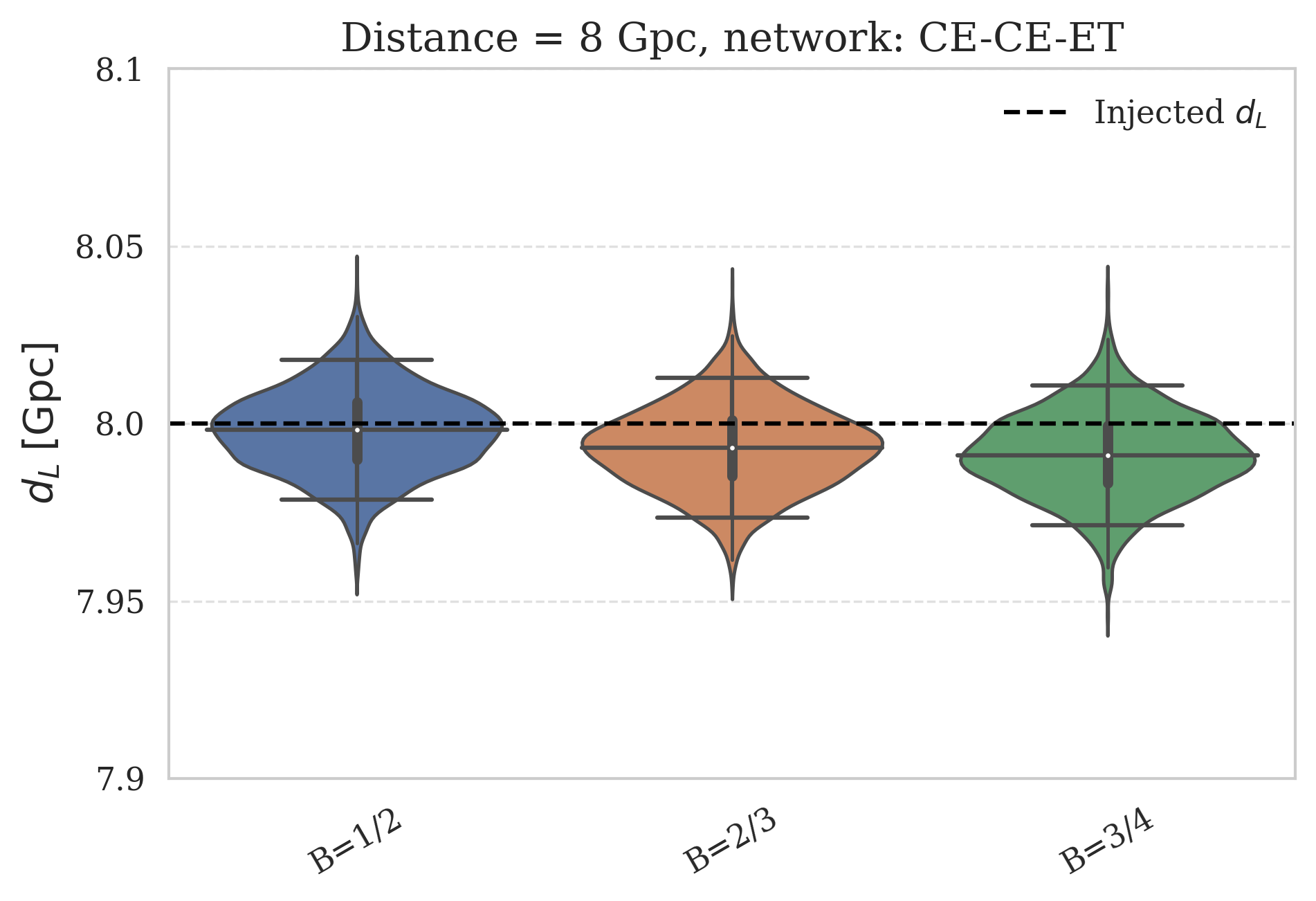}
    \end{minipage}\hfill
    \begin{minipage}{0.3\textwidth}
        \centering
        \includegraphics[width=\textwidth]{paper-figures/distance_10Gpc/dL_vs_B_distance_10000_network_L1_CE_H1_CE_V1_ET.png}
    \end{minipage}

    \vspace{0.5em}

     \begin{minipage}{0.3\textwidth}
        \centering
        \includegraphics[width=\textwidth]{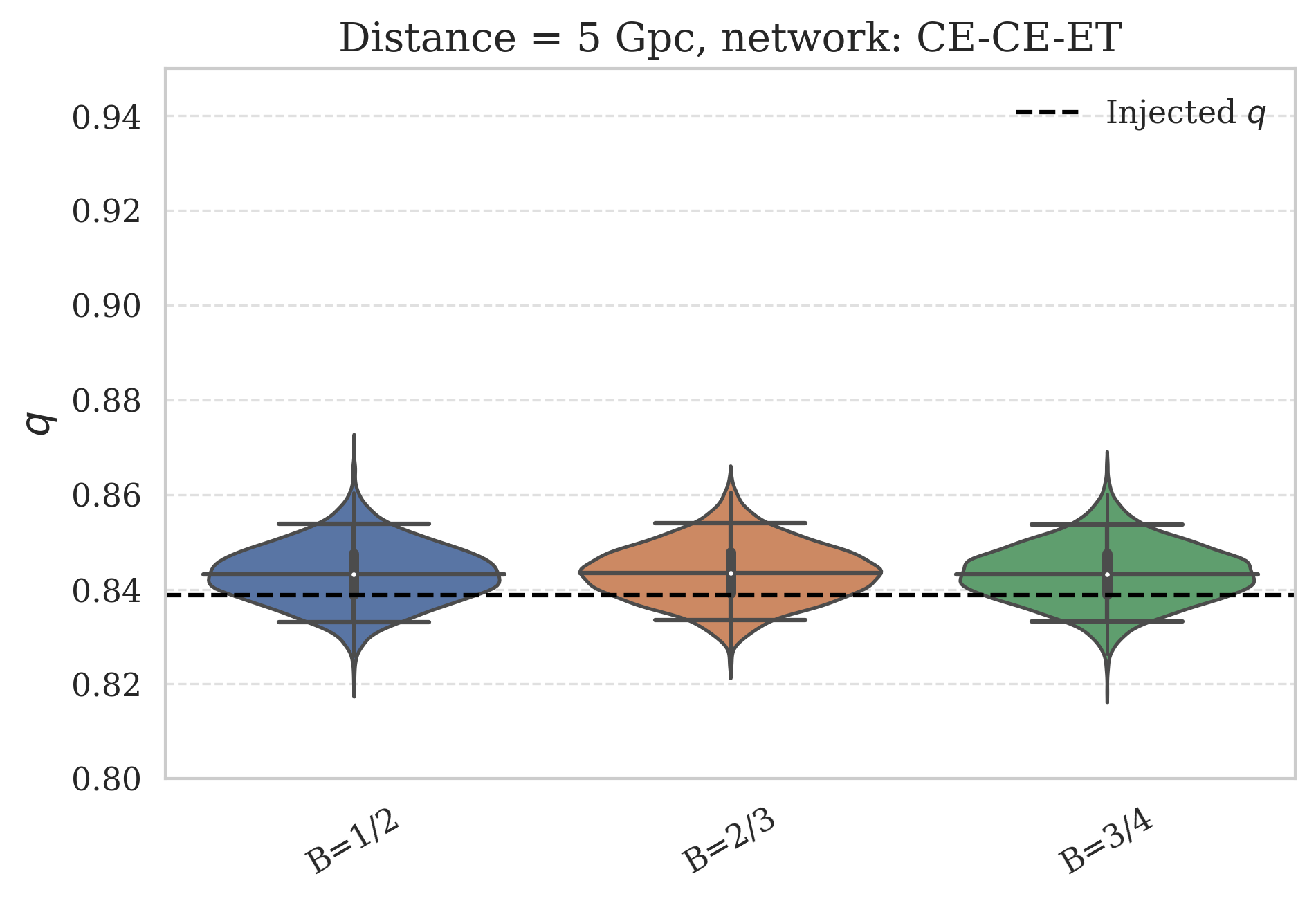}
    \end{minipage}\hfill
    \begin{minipage}{0.3\textwidth}
        \centering
        \includegraphics[width=\textwidth]{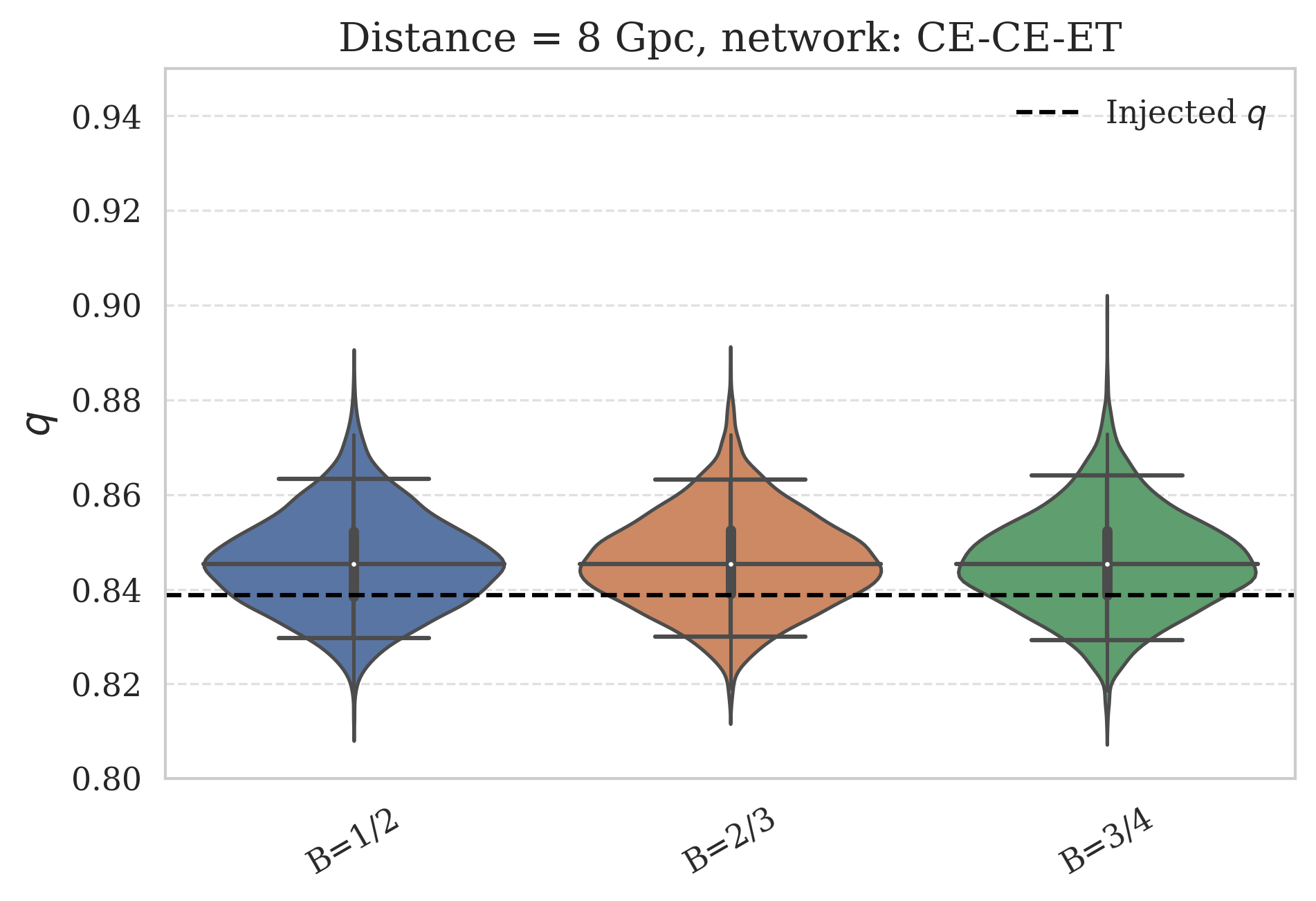}
    \end{minipage}\hfill
    \begin{minipage}{0.3\textwidth}
        \centering
        \includegraphics[width=\textwidth]{paper-figures/distance_10Gpc/q_vs_B_distance_10000_network_L1_CE_H1_CE_V1_ET.png}
    \end{minipage}

    \vspace{0.5em}

    \begin{minipage}{0.3\textwidth}
        \centering
        \includegraphics[width=\textwidth]{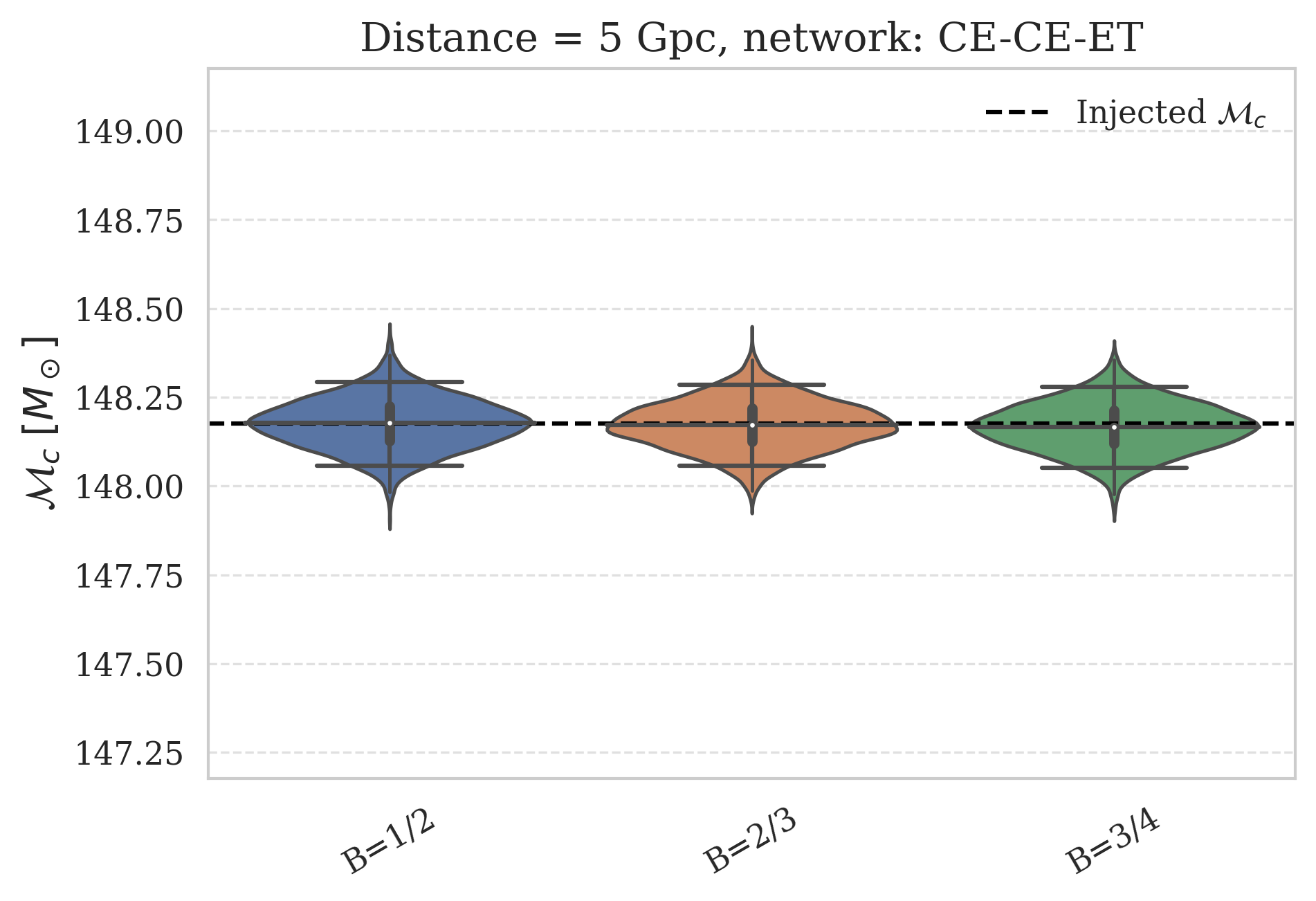}
    \end{minipage}\hfill
    \begin{minipage}{0.3\textwidth}
        \centering
        \includegraphics[width=\textwidth]{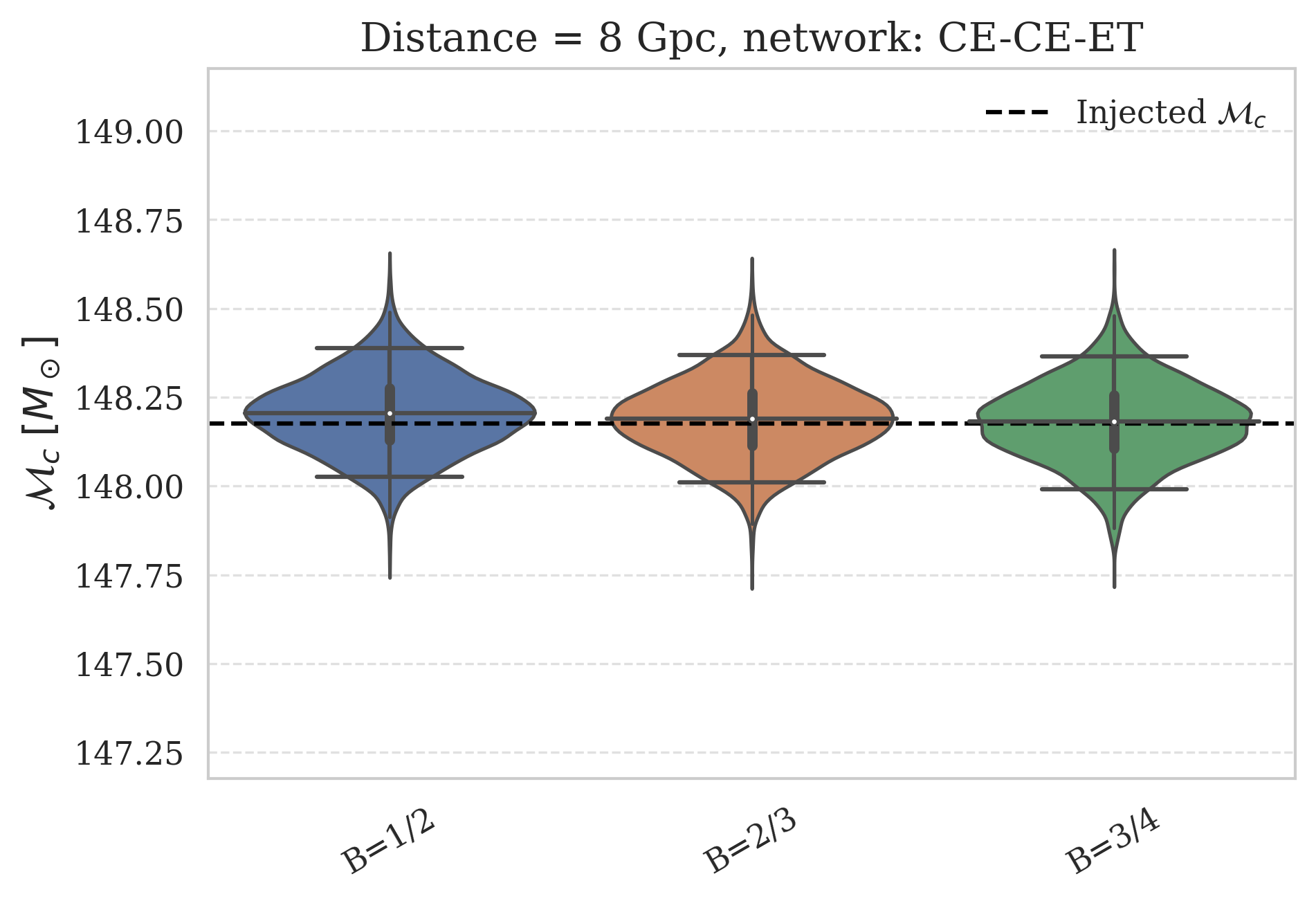}
    \end{minipage}\hfill
    \begin{minipage}{0.3\textwidth}
        \centering
        \includegraphics[width=\textwidth]{paper-figures/distance_10Gpc/Mc_vs_B_distance_10000_network_L1_CE_H1_CE_V1_ET.png}
    \end{minipage}

    \caption{
       CE-CE-ET network: Posterior distributions of luminosity distance $d_L$ (top row), mass ratio $q$ (middle row), and chirp mass $\mathcal{M}_c$ (bottom row) for signals injected at different distances.}
    \label{fig:3x3_fullpage_posteriors_cbc_parameters_ce_ce_et}
\end{figure*}

\section{ICM with CE-CE-ET and CE-CE-LI networks at multiple distances and choice of sky location}

\subsection{ICM with next-generation networks}
In the main letter, we presented the posterior distributions of the chirp mass $\mathcal{M}_c$, mass ratio $q$, and luminosity distance $d_L$ for a signal injected at 10 Gpc for the CE-CE-ET configuration. In this section, we show the posteriors for all three injected distances, that is - 5 Gpc, 8 Gpc and 10 Gpc, and for two detector configurations: the CE–CE–ET and CE–CE–LI networks. 

The subplots in Fig.~\ref{fig:3x3_fullpage_posteriors_cbc_parameters_ce_ce_et} show the violin plots for the posterior distribution of the parameters of the injected CBC signal. The injected signal falls within the 90\% credible interval for all the parameters for the CE–CE–ET network. This shows the robustness of the sampler even in the presence of the gravitational-wave memory signal. As expected, the posterior widths increase with distance due to the reduction in signal-to-noise ratio (SNR), leading to larger uncertainties in parameter estimation. Although the memory contribution does introduce non-zero frequency components near the merger, its effect on the parameter estimation of the CBC signal is negligible over the range of distances considered.

\begin{figure*}[htb!]
    \centering

    \begin{minipage}{0.3\textwidth}
        \centering
        \includegraphics[width=\textwidth]{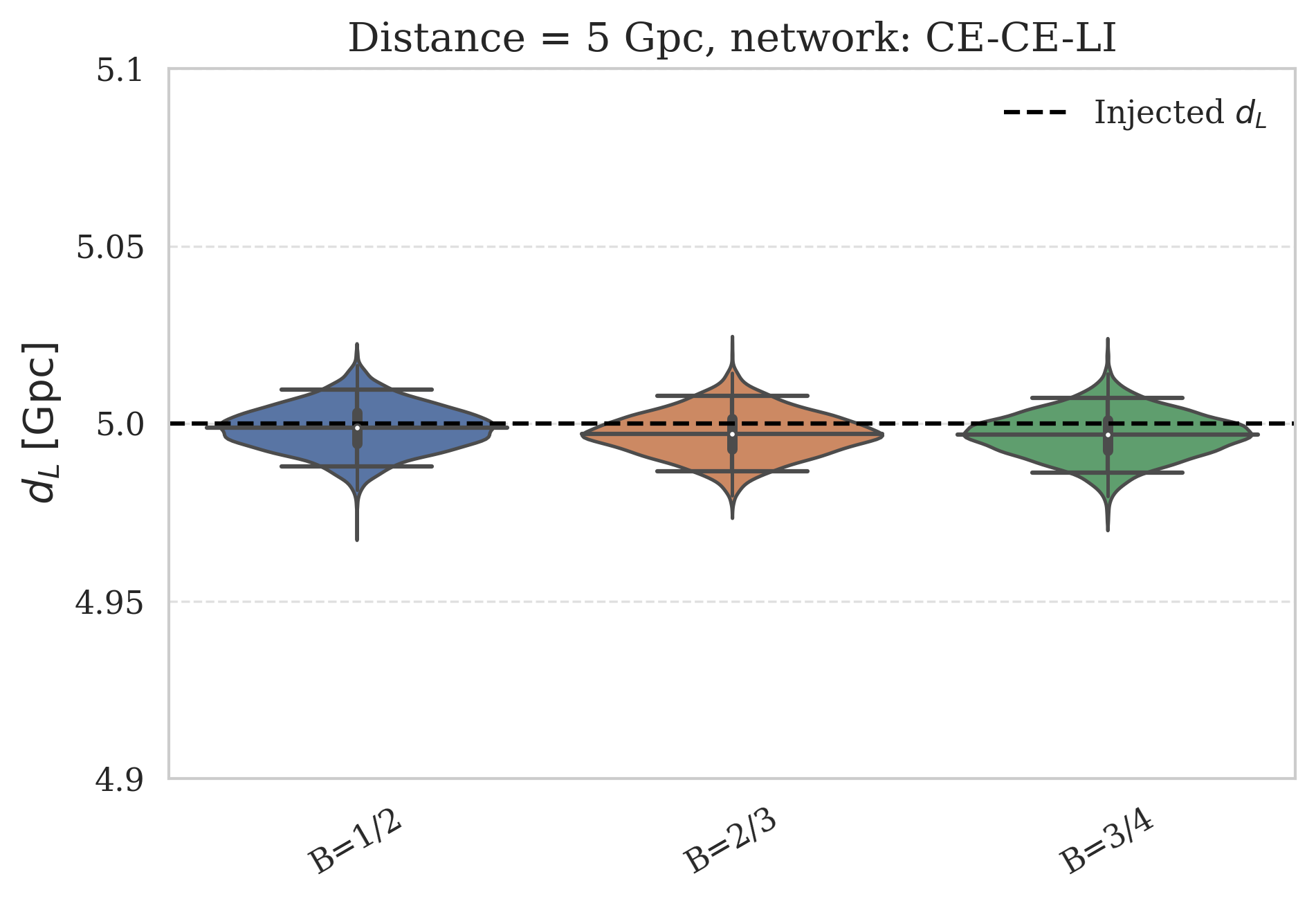}
    \end{minipage}\hfill
    \begin{minipage}{0.3\textwidth}
        \centering
        \includegraphics[width=\textwidth]{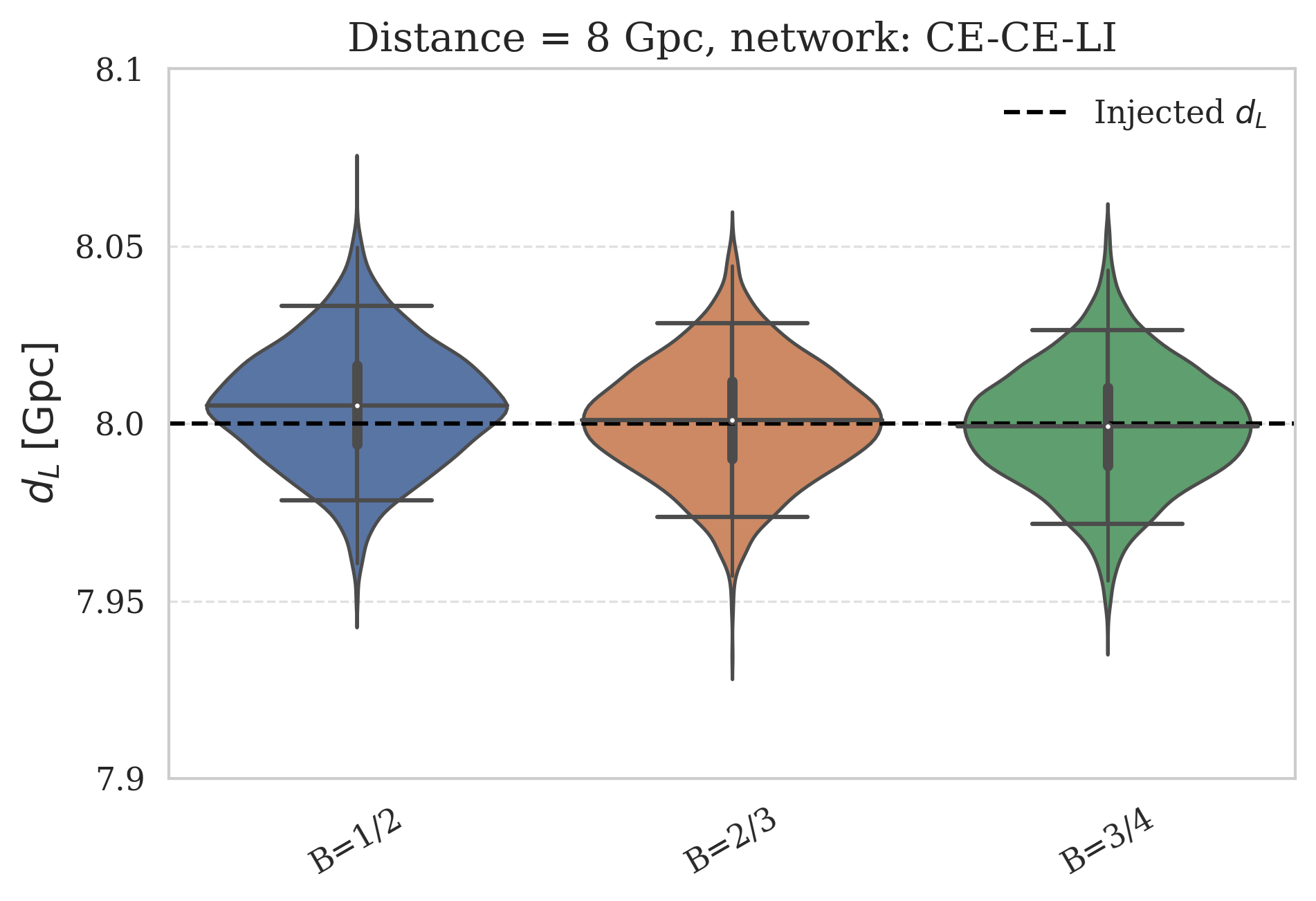}
    \end{minipage}\hfill
    \begin{minipage}{0.3\textwidth}
        \centering
        \includegraphics[width=\textwidth]{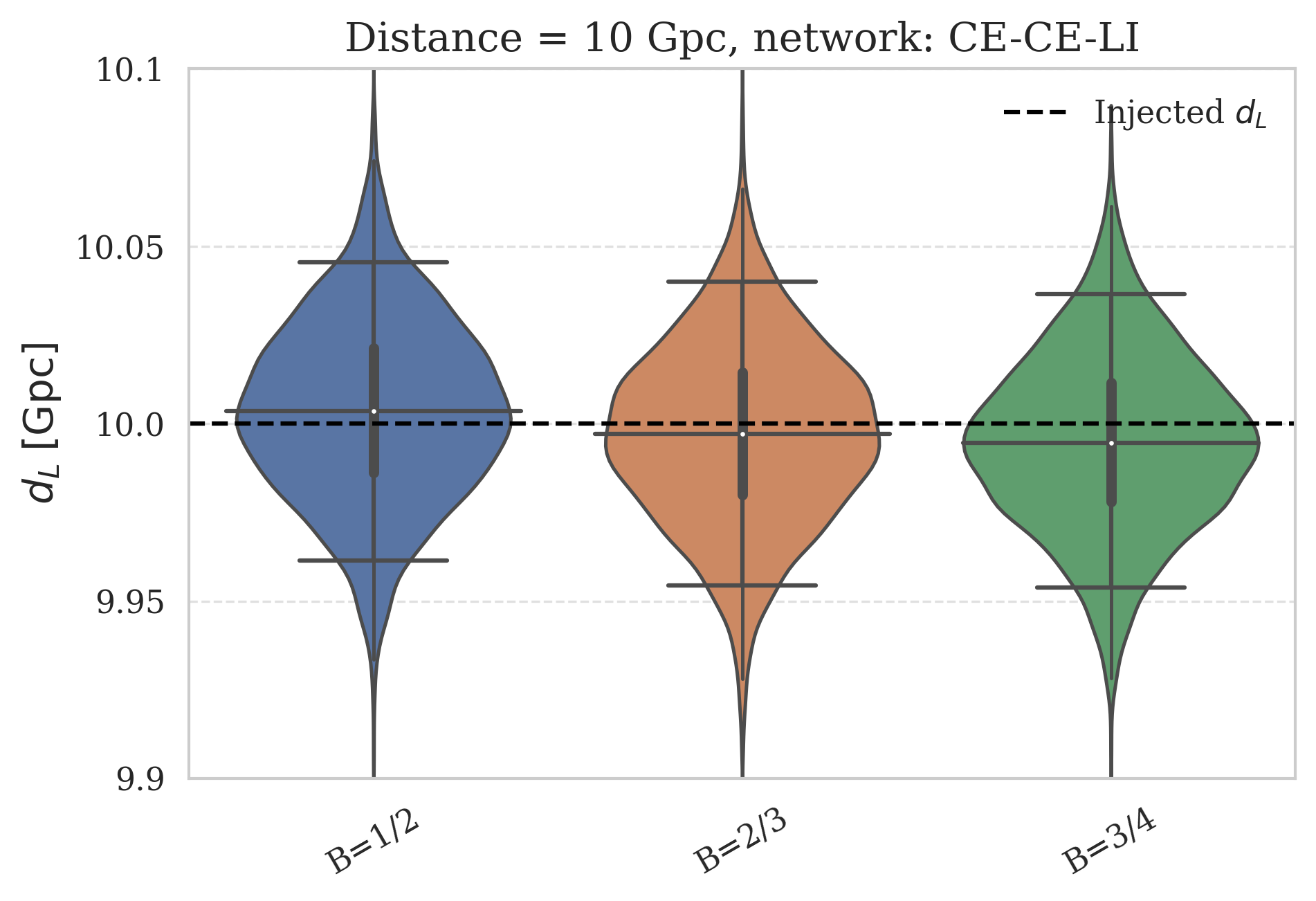}
    \end{minipage}

    \vspace{0.5em}

    \begin{minipage}{0.3\textwidth}
        \centering
        \includegraphics[width=\textwidth]{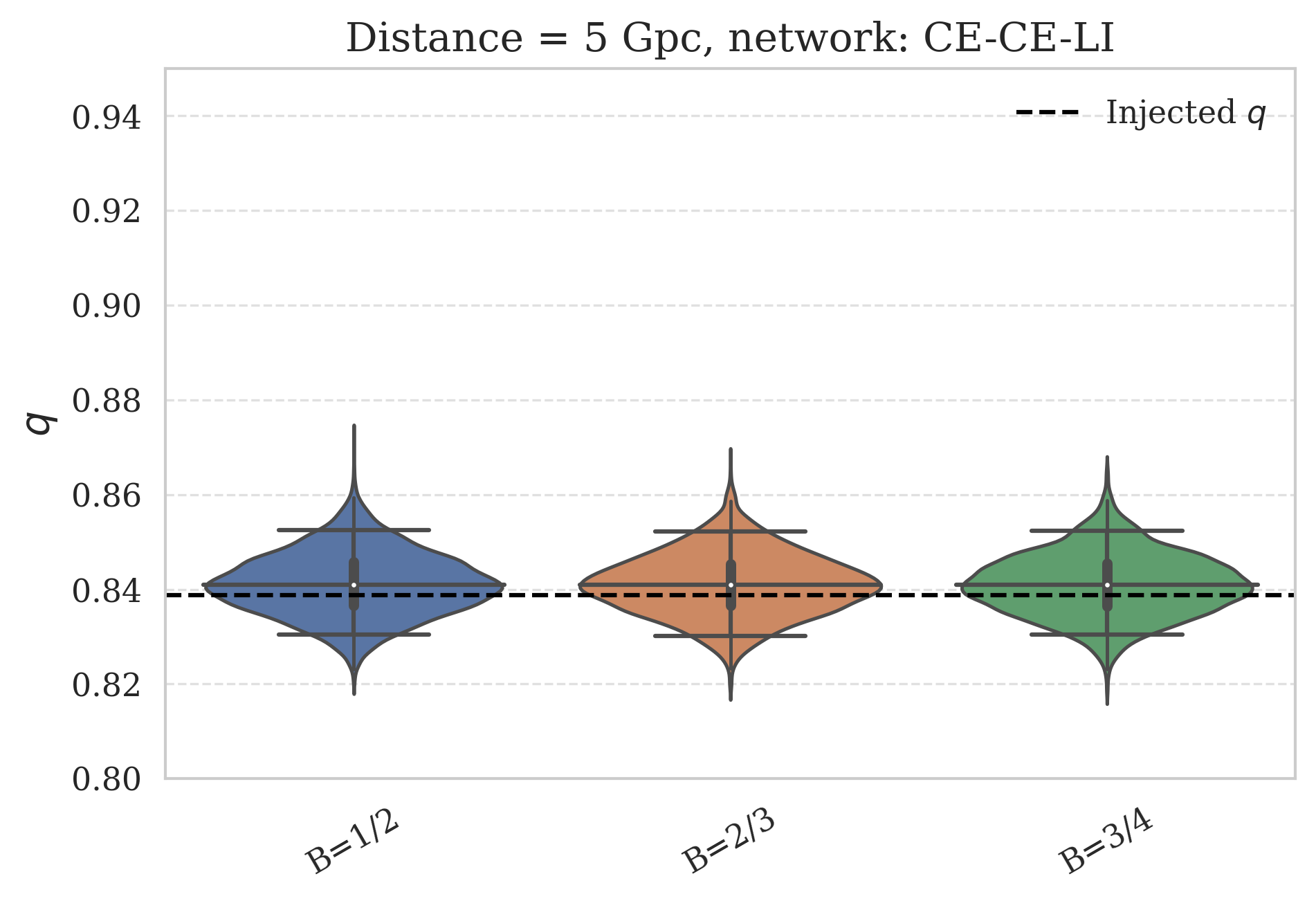}
    \end{minipage}\hfill
    \begin{minipage}{0.3\textwidth}
        \centering
        \includegraphics[width=\textwidth]{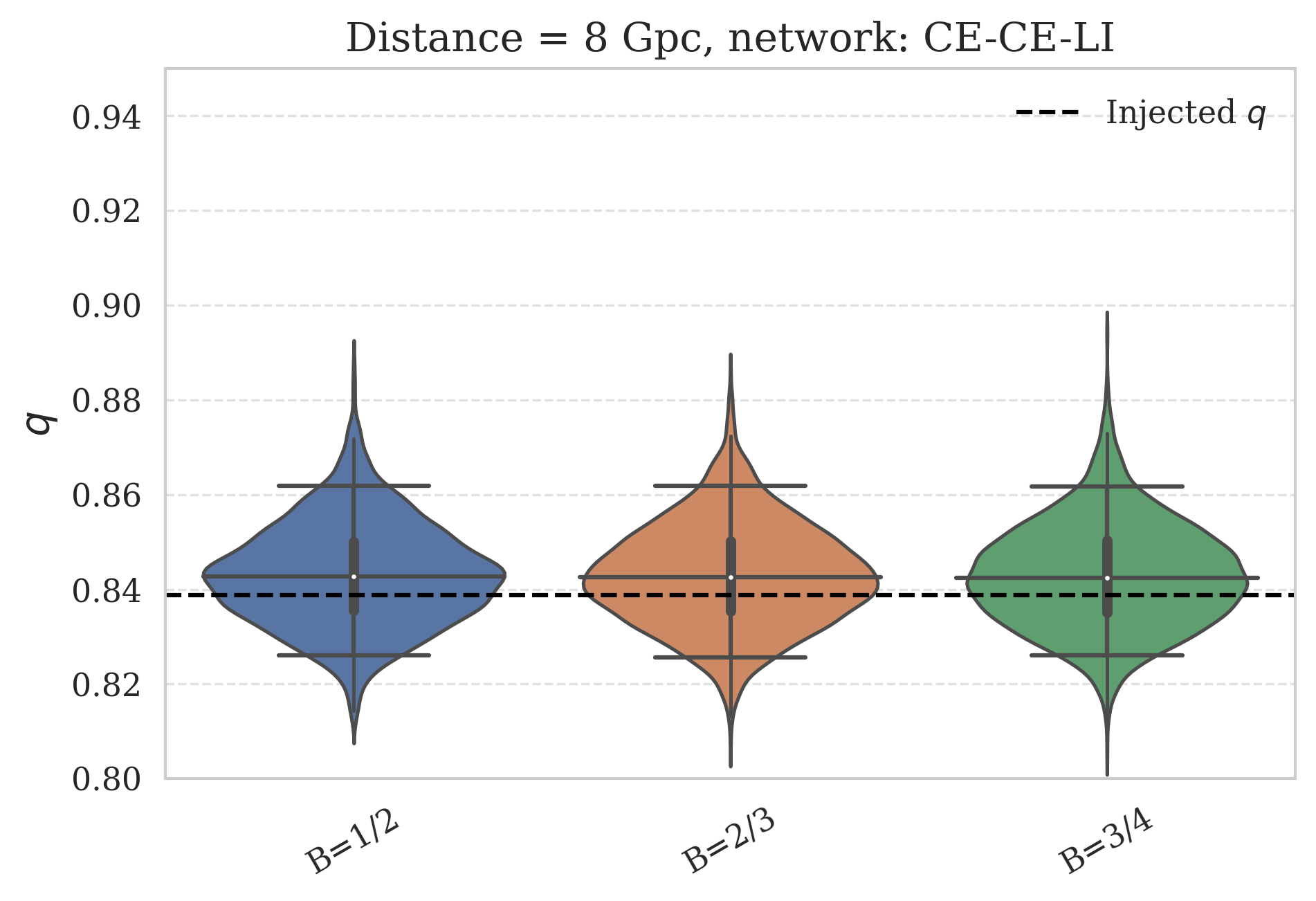}
    \end{minipage}\hfill
    \begin{minipage}{0.3\textwidth}
        \centering
        \includegraphics[width=\textwidth]{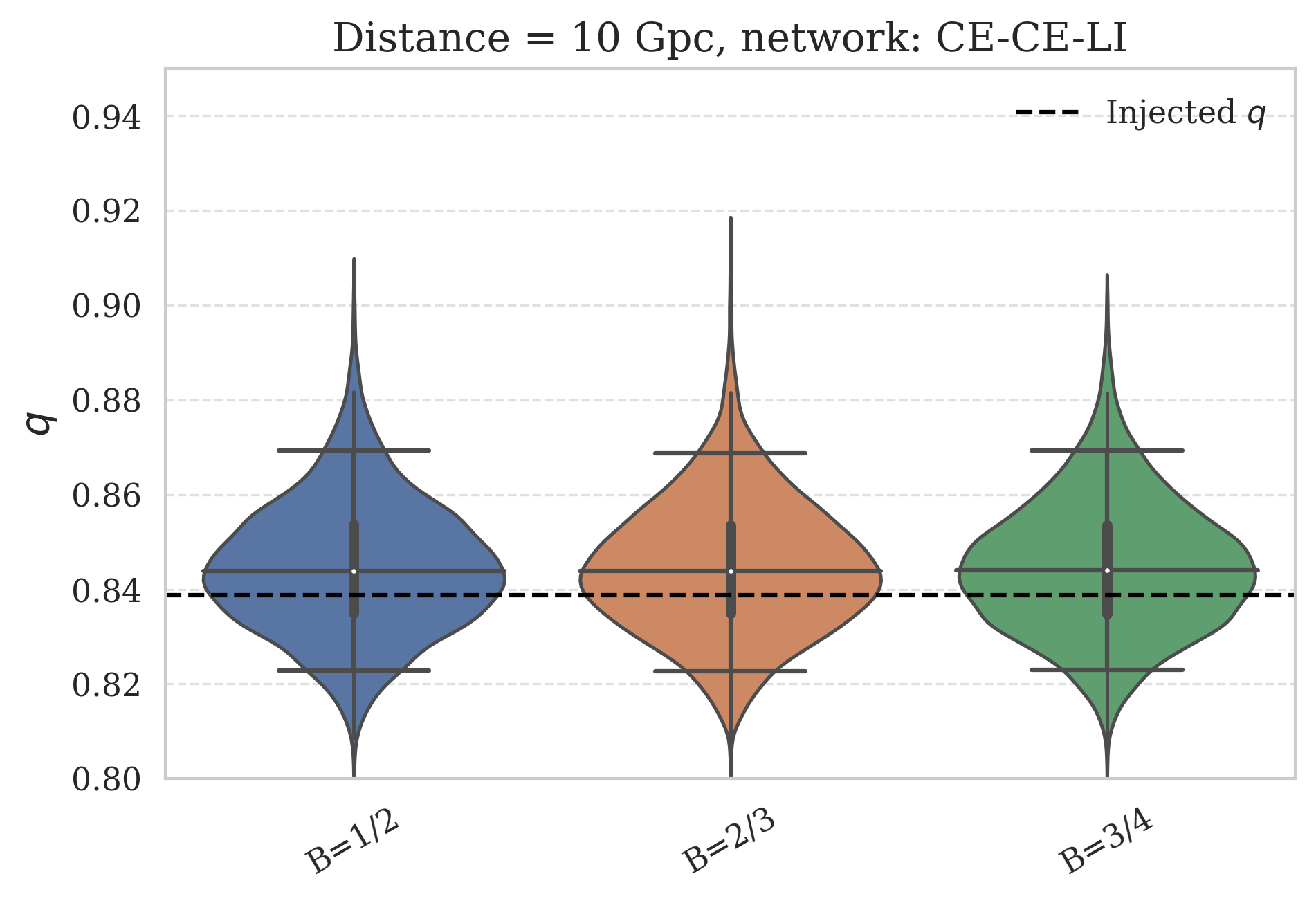}
    \end{minipage}

    \vspace{0.5em}

    \begin{minipage}{0.3\textwidth}
        \centering
        \includegraphics[width=\textwidth]{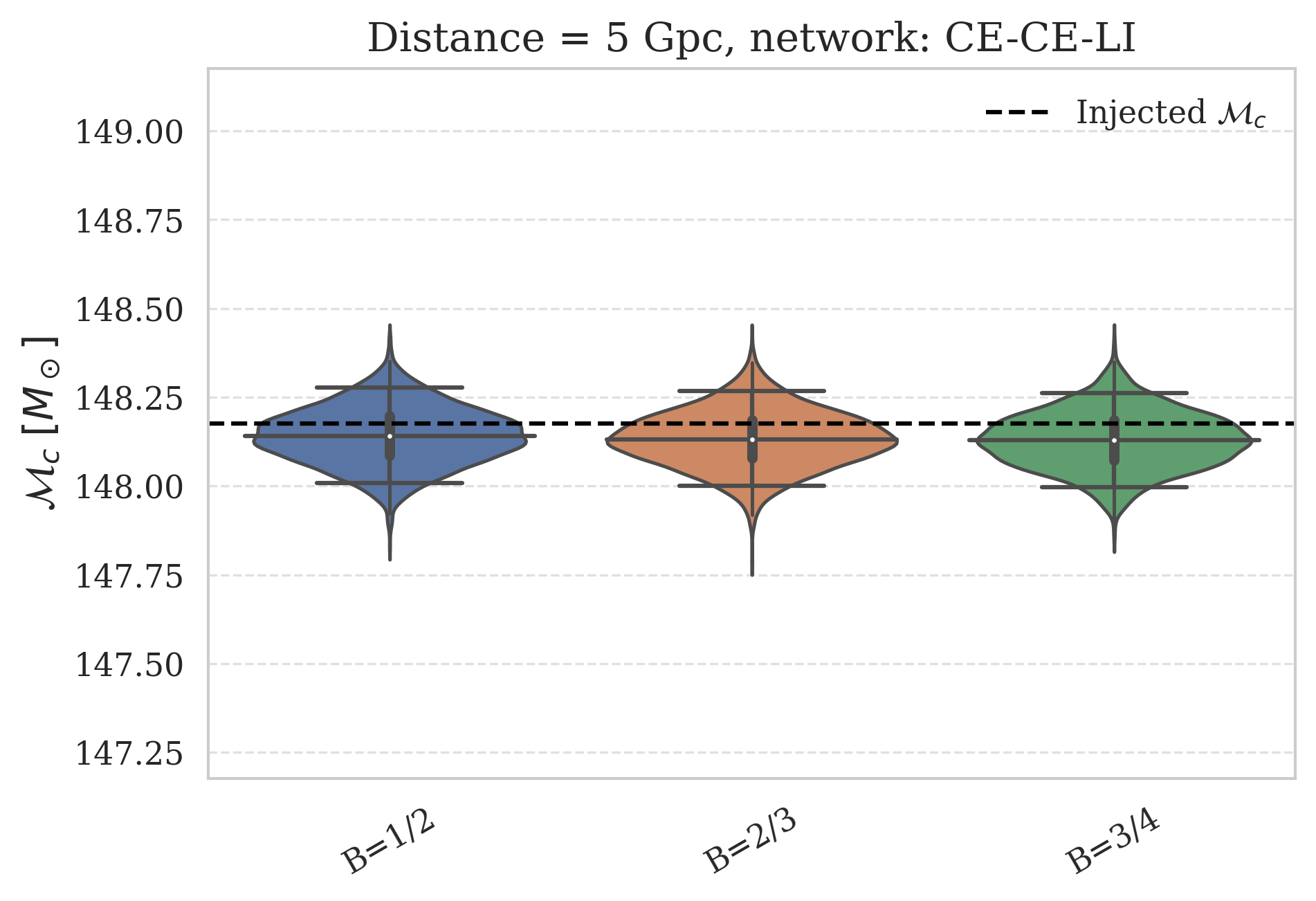}
    \end{minipage}\hfill
    \begin{minipage}{0.3\textwidth}
        \centering
        \includegraphics[width=\textwidth]{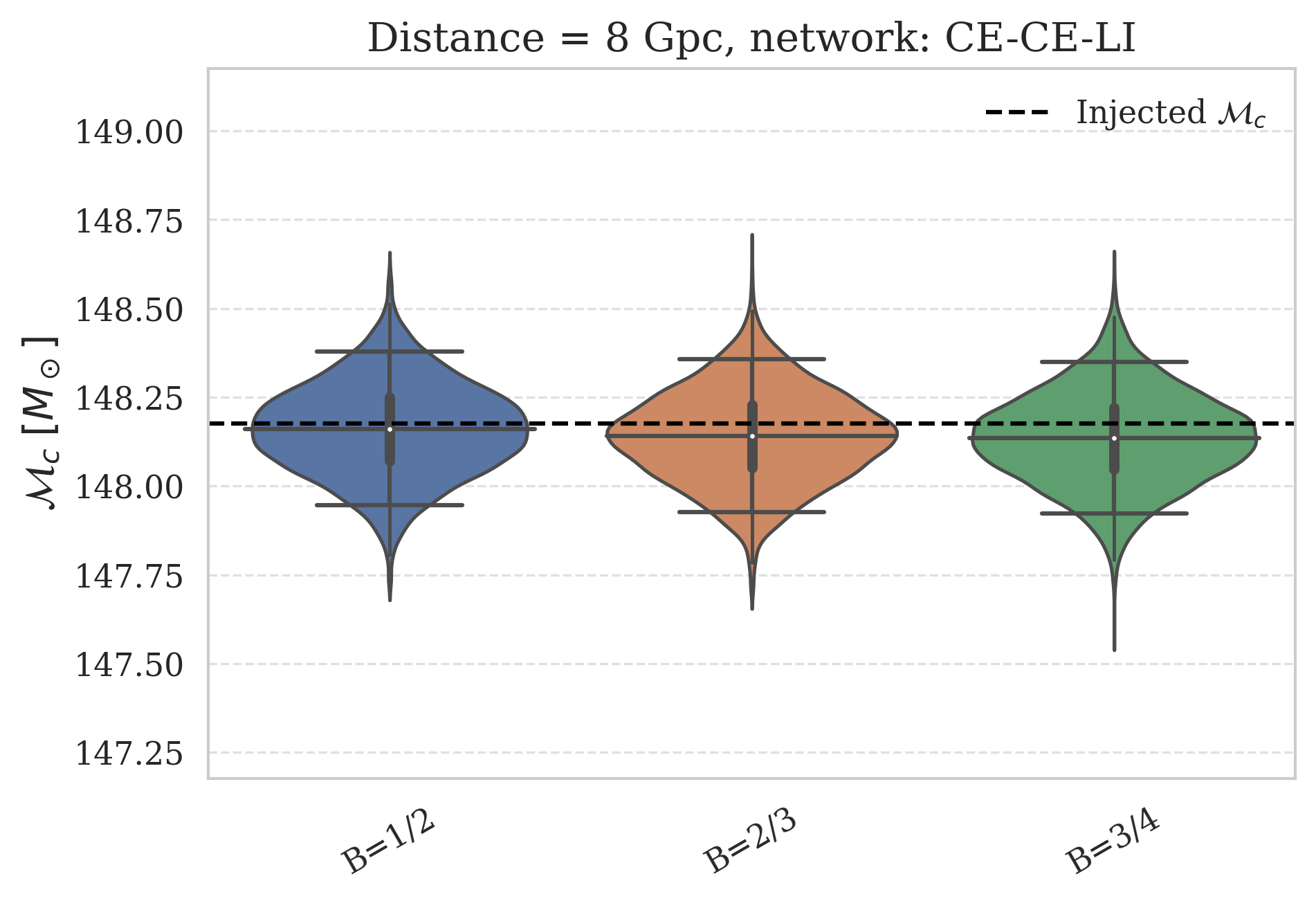}
    \end{minipage}\hfill
    \begin{minipage}{0.3\textwidth}
        \centering
        \includegraphics[width=\textwidth]{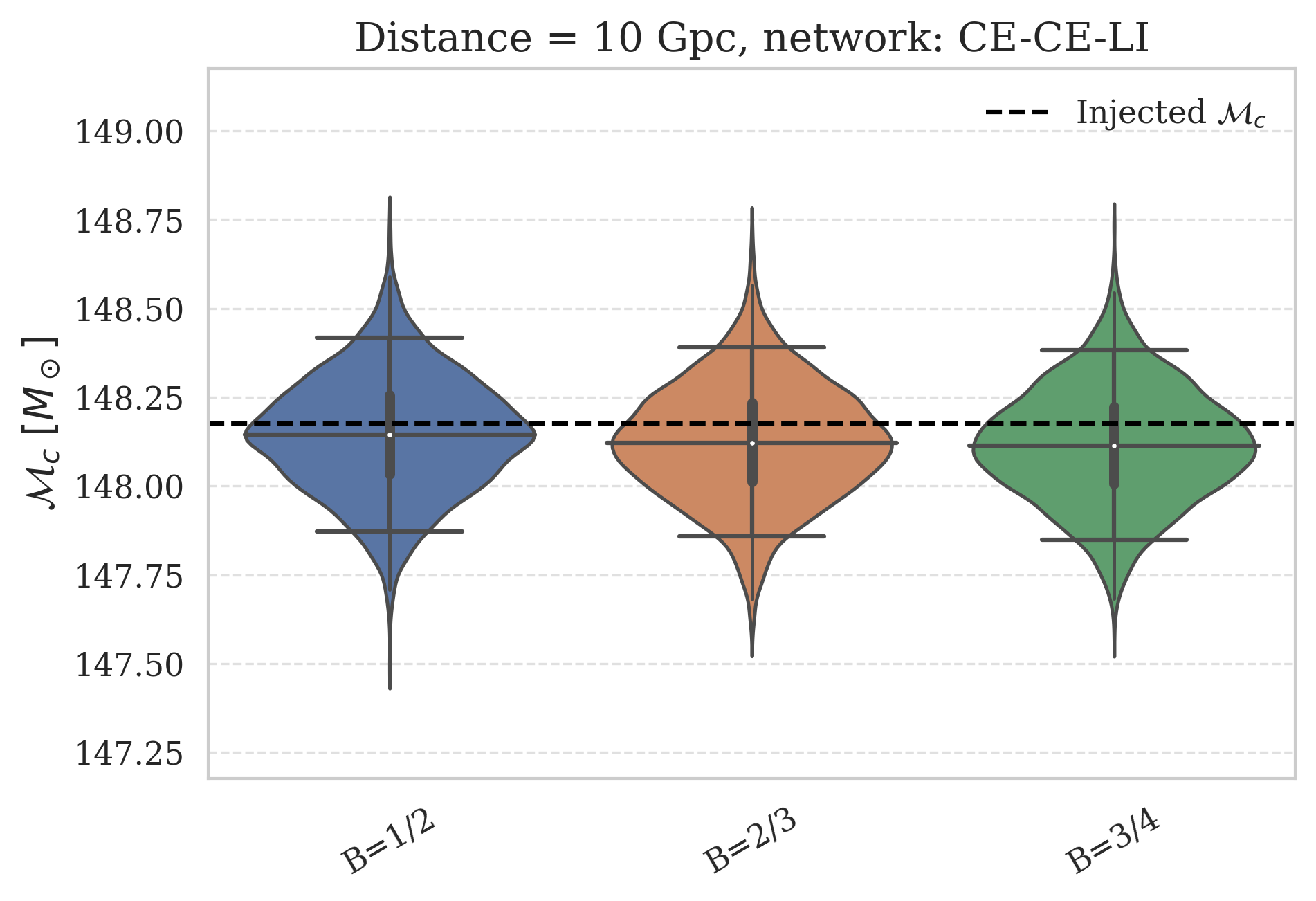}
    \end{minipage}
    \caption{
       CE-CE-LI network: Posterior distributions of luminosity distance $d_L$ (top row), mass ratio $q$(middle row) and chirp mass $\mathcal{M}_c$ (bottom row). The posteriors shown in the left, middle and right columns are for the signals injected at $d_L = 5\, \mathrm{Gpc}, 8\, \mathrm{Gpc}, 10 \,\mathrm{Gpc}$ respectively. Each figure shows the posteriors in the form of violin plots for the three cosmologies considered in this work - $B = \frac{1}{2}$, $\frac{2}{3}$ and $\frac{3}{4}$. We see that the estimates of all three parameters are both precise and accurate.}  \label{fig:3x3_fullpage_posteriors_cbc_parameters_ce_ce_aplus}
\end{figure*}

\begin{figure*}[h!]
    \centering
    \includegraphics[width=0.49\linewidth]{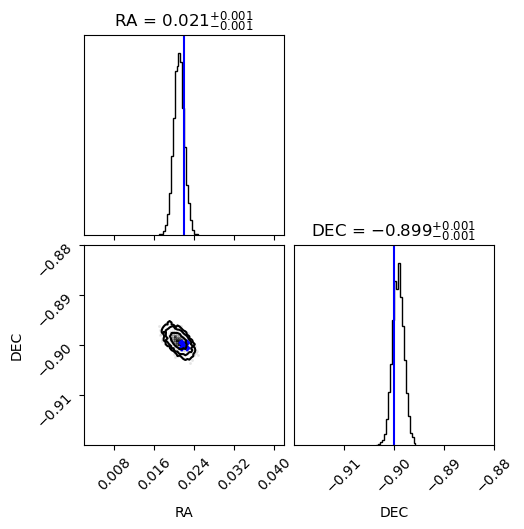}
    \includegraphics[width=0.49\linewidth]{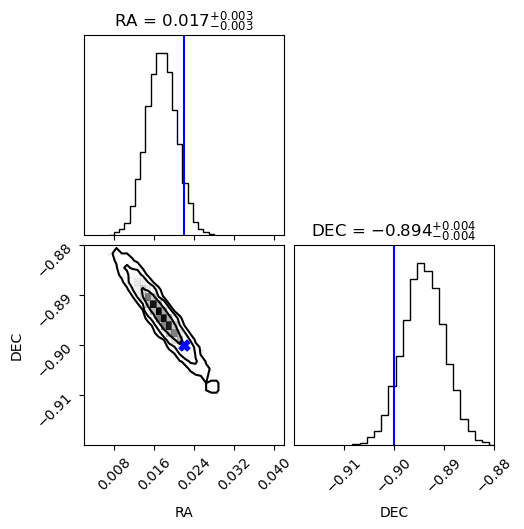}
        \caption{Plots of the sky location posteriors. The top panel shows the sky location posteriors for the CE-CE-ET network, and the bottom panel shows the sky location posteriors for the CE-CE-LI network, for $B = \frac{1}{2}$ at $d_L = 5$GPc. The blue cross shows the injected right ascension and declination values on the 2D contour, and the vertical blue lines show these same values overlaid with the histograms. The sky location posterior is precise and contains the injected values near its centre.}
    \label{fig:skyloc}
\end{figure*}

The recovered constraints on the cosmological parameter $B$ for the CE–CE–ET network are summarized in Table \ref{tab:B_constraints_ceceet} (for $H_0 = 70$ km/s/Mpc), Table \ref{tab:B_constraints_ceceet_H0_73} (for $H_0 = 73$ km/s/Mpc), and Table \ref{tab:B_constraints_ceceet_H0_67} (for $H_0=67$ km/s/Mpc). These three median estimates of the Hubble parameter are taken from the BNS measurements \citep{Abbott:2017-H0}, the SH0ES measurements \citep{shoes-H0} and the Planck measurements \citep{planck-H0} respectively. Since there is currently no consensus on the value of $H_0$, with different methods yielding different values, we tested our method with $H_0$ values estimated from these three methods. We observe that while the $B$ estimates are not completely independent of $H_0$, the variation from one $H_0$ value to another is minimal - the median values varying typically by $\sim 2-3\%$ and the error bars varying typically by $\sim 5-15\%$. 

\begin{table}[!htb]
\caption{\label{tab:B_constraints_ceceet}
Recovered $90\%$ credible intervals for $B$ using the CE-CE-ET network and $H_0 = 70$ km/s/Mpc}
\centering
\begin{tabular*}{\textwidth}{@{\extracolsep\fill}cccc}
\hline \hline
\textbf{Distance} & \multicolumn{3}{c}{\textbf{Recovered $B$ (90\% CI)}} \\
(Gpc) & $B_{\rm inj}=0.50$ & $B_{\rm inj}=0.67$ & $B_{\rm inj}=0.75$ \\
\hline

5 &
$0.499^{+0.064}_{-0.049}$ &
$0.703^{+0.116}_{-0.116}$ &
$0.827^{+0.118}_{-0.177}$ \\[4pt]

\hline

8 &
$0.497^{+0.062}_{-0.047}$ &
$0.675^{+0.087}_{-0.085}$ &
$0.770^{+0.123}_{-0.119}$ \\[4pt]

\hline

10 &
$0.496^{+0.062}_{-0.050}$ &
$0.670^{+0.083}_{-0.080}$ &
$0.757^{+0.116}_{-0.090}$ \\[4pt]

\hline \hline
\end{tabular*}
\end{table}

\begin{table}[!htb]
\caption{\label{tab:B_constraints_ceceet_H0_73}
Recovered $90\%$ credible intervals for $B$ using the CE-CE-ET network and $H_0 = 73$ km/s/Mpc}
\centering
\begin{tabular*}{\textwidth}{@{\extracolsep\fill}cccc}
\hline \hline
 \textbf{Distance} & \multicolumn{3}{c}{\textbf{Recovered $B$ (90\% CI)}} \\
(Gpc) & $B_{\rm inj}=0.50$ & $B_{\rm inj}=0.67$ & $B_{\rm inj}=0.75$ \\
\hline

5 &
$0.499^{+0.061}_{-0.045}$ &
$0.691^{+0.095}_{-0.105}$ &
$0.800^{+0.126}_{-0.150}$ \\[4pt]

\hline

8 &
$0.496^{+0.056}_{-0.044}$ &
$0.662^{+0.090}_{-0.064}$ &
$0.765^{+0.110}_{-0.096}$ \\[4pt]

\hline

10 &
$0.497^{+0.062}_{-0.047}$ &
$0.665^{+0.088}_{-0.068}$ &
$0.758^{+0.110}_{-0.084}$ \\[4pt]

\hline \hline
\end{tabular*}
\end{table}

\begin{table}[h]
\caption{\label{tab:B_constraints_ceceet_H0_67}
Recovered $90\%$ credible intervals for $B$ using the CE-CE-ET network and $H_0 = 67$ km/s/Mpc}
\centering
\begin{tabular*}{\textwidth}{@{\extracolsep\fill}cccc}
\hline \hline
 \textbf{Distance} & \multicolumn{3}{c}{\textbf{Recovered $B$ (90\% CI)}} \\
(Gpc) & $B_{\rm inj}=0.50$ & $B_{\rm inj}=0.67$ & $B_{\rm inj}=0.75$ \\
\hline

5 &
$0.499^{+0.074}_{-0.053}$ &
$0.691^{+0.132}_{-0.105}$ &
$0.821^{+0.149}_{-0.181}$ \\[4pt]

\hline

8 &
$0.497^{+0.062}_{-0.047}$ &
$0.687^{+0.085}_{-0.101}$ &
$0.776^{+0.125}_{-0.125}$ \\[4pt]

\hline

10 &
$0.496^{+0.064}_{-0.051}$ &
$0.665^{+0.090}_{-0.074}$ &
$0.772^{+0.109}_{-0.111}$ \\[4pt]
\hline \hline
\end{tabular*}
\end{table}

We now extend our analysis to the CE–CE–LI network, where LIGO-India (LI) operates at Aplus sensitivity. For LI, we adopt the analytical power spectral density (PSD) corresponding to \texttt{aLIGOAPlusDesignSensitivityT1800042} \citep{alex_nitz_2024_10473621}, as implemented in \texttt{PyCBC} \citep{alex_nitz_2024_10473621}. This configuration complements the CE detectors while ensuring that at least two instruments have sufficiently high memory SNR for robust parameter inference. Using the same set of signal injections as in the CE–CE–ET case, we perform parameter estimation and infer the cosmological parameter $B$. The posteriors are depicted in Fig. \ref{fig:3x3_fullpage_posteriors_cbc_parameters_ce_ce_aplus} and the recovered $B$ constraints are presented in Table ~\ref{tab:B_constraints_ceceli}. We find that the injected values of the CBC parameters are consistently recovered within the $90\%$ credible intervals across all distances considered (similar to the CE-CE-ET network), although the CE–CE–LI configuration yields slightly broader posteriors, reflecting the comparatively lower sensitivity of LIGO-India. The $B$ constraints show similar trends as the CE-CE-ET network,  with similar uncertainties. 

\begin{table}[!htb]
\caption{\label{tab:B_constraints_ceceli}
Recovered $90\%$ credible intervals for $B$ using the CE-CE-LI network ($H_0 = 70$ km/s/Mpc).}
\centering
\begin{tabular*}{\textwidth}{@{\extracolsep\fill}cccc}
\hline \hline
 \textbf{Distance} & \multicolumn{3}{c}{\textbf{Recovered $B$ (90\% CI)}} \\
(Gpc) & $B_{\rm inj}=0.50$ & $B_{\rm inj}=0.67$ & $B_{\rm inj}=0.75$ \\
\hline

5  &
$0.501^{+0.074}_{-0.055}$ &
$0.697^{+0.132}_{-0.111}$ &
$0.821^{+0.153}_{-0.181}$ \\[4pt]

\hline

8  &
$0.497^{+0.066}_{-0.051}$ &
$0.671^{+0.106}_{-0.081}$ &
$0.815^{+0.107}_{-0.165}$ \\[4pt]

\hline

10 &
$0.496^{+0.078}_{-0.060}$ &
$0.675^{+0.099}_{-0.084}$ &
$0.799^{+0.097}_{-0.150}$ \\[4pt]
\hline \hline
\end{tabular*}
\end{table}

Comparing the tables between \ref{tab:B_constraints_ceceet} and \ref{tab:B_constraints_ceceli}, we observe a  preference for the CE–CE–ET configuration which manages to place slightly better constraints on $B$. 

\begin{figure*}[!htb]
    \centering
    \includegraphics[width=\linewidth]{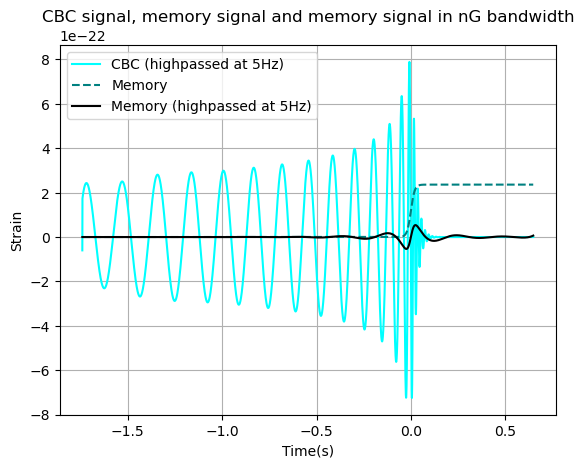}
    \caption{The CBC signal at distance = 5 Gpc in the time domain overlaid on the ICM signal corresponding to $B = \frac{1}{2}$. The actual memory signal, as predicted by the GR is a step-like function (cyan dashed). The signal observed in the detector noise will be high-passed version (shown in the solid black). This figure illustrates the faintness of the memory signal in the detector bandwidth compared to the CBC signal.}
    \label{fig:cbc_vs_memory}
\end{figure*}

\subsection{Choice of sky location}
\begin{figure*}
    \includegraphics[width=0.49\textwidth]{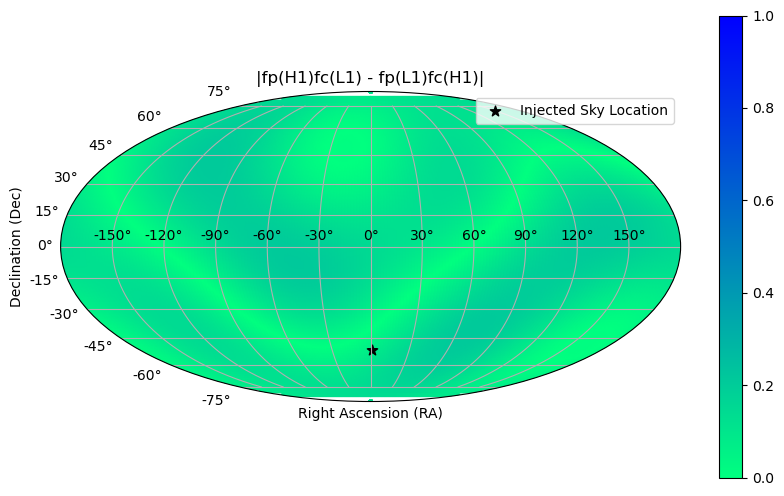}
    \includegraphics[width=0.49\textwidth]{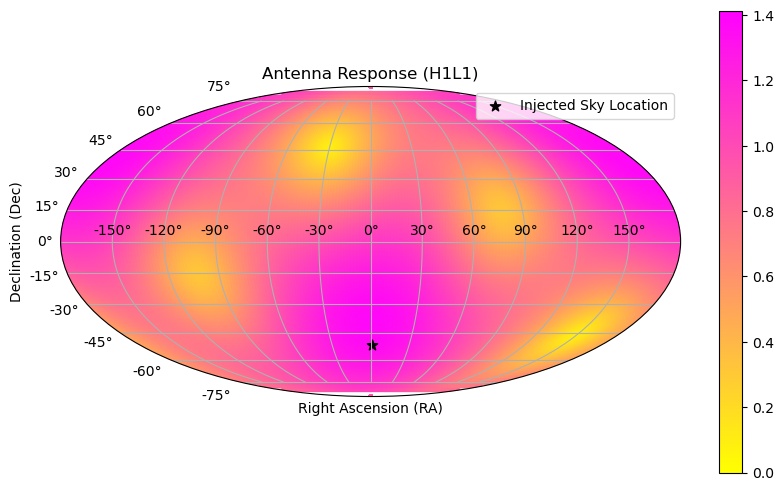}
    \includegraphics[width=0.49\textwidth]{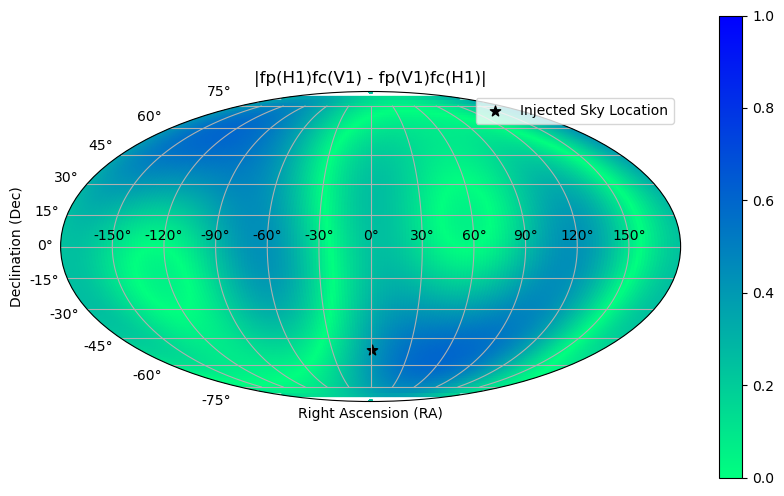}
    \includegraphics[width=0.49\textwidth]{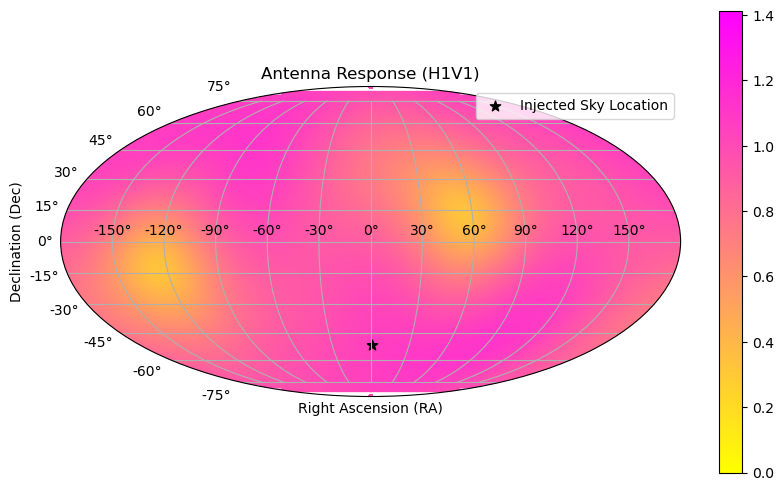}
    \caption{The left column shows the variations of the determinant of the F-matrix for H1 and L1 detectors (upper panel) and H1 and V1 detectors (lower panel) across the entire sky. The right panel shows the variations of the antenna response ($\sqrt{f_p^2 + f_c^2}$) across the entire sky for the H1-L1 network (upper panel) and H1-V1 network (lower panel). The black star shows the sky location chosen for this work. In the left column, we see that the injected sky location lies in a region which has a tolerably high F-matrix determinant (around 0.1 for the H1L1 network, which is sufficiently high for meaningful reconstruction of the plus and cross polarization components) and in the right column we see that it lies in the pink zone, that is, the zone of high antenna response.}
    \label{fig:sky_location}
\end{figure*}

As shown in Fig. \ref{fig:sky_location}, the chosen sky location has two advantages (though this is by no means the only location with these advantages) - the $F$-matrix (described in the letter) is invertible and the antenna response is high. While an invertible F-matrix makes it possible to resolve the polarization components of the memory signal (a crucial stage of the method), the high antenna response ensures that the SNR of the memory signal is high enough for detection and characterization. Fig. \ref{fig:sky_location} also shows us that there are lots of other sky locations which boast F-matrix invertibility and high antenna response, indicating that our method is applicable to large patches of the sky, rather than a few isolated regions. 

\section{Matched-filter method to estimate memory offset}

\begin{figure*}[!hbt]
    \centering
    \includegraphics[width=0.45\textwidth]{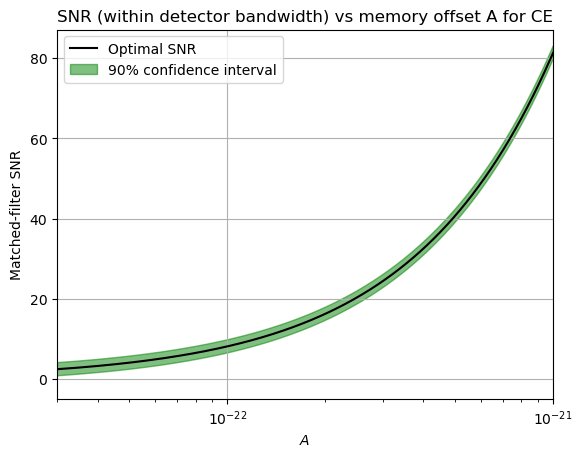}
    \includegraphics[width=0.45\textwidth]{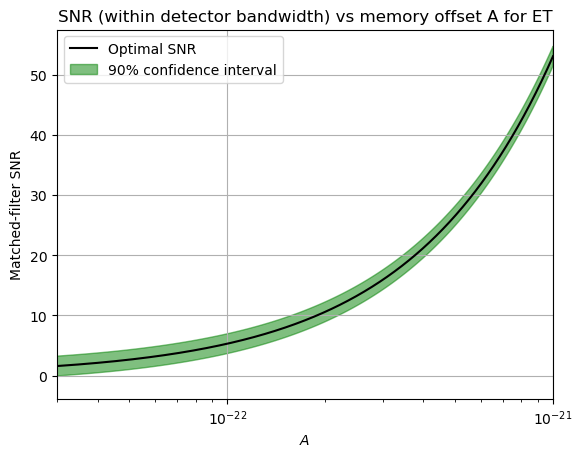}
    \caption{Plots used to read off memory offset from the estimated matched-filter SNR with the memory template. The black solid curve shows the variation of the optimal SNR with the memory offset. The shaded green region shows the $90\%$ confidence region around the optimal SNR. For a measured SNR on the y-axis, the corresponding $A$ is not a single number, but rather an interval inferred from these plots.}
    \label{fig:bookkeeping}
\end{figure*}

This section details the matched-filter based method to estimate the memory offset from the strain data, utilizing the distinct morphological differences between the astrophysical CBC signal and the memory transient (see Fig. \eqref{fig:cbc_vs_memory}).

As shown in Figs. \ref{fig:3x3_fullpage_posteriors_cbc_parameters_ce_ce_et}, \ref{fig:3x3_fullpage_posteriors_cbc_parameters_ce_ce_aplus} and \ref{fig:skyloc}, the astrophysical parameters of a CBC signal are estimated to a fair degree of accuracy when the recovery waveform model is the same as that of the injected waveform model in a simulation study. However, in a real scenario, there will not be any injected model, but rather a {\it true} signal present in the data and we have to develop waveform models which describe the true signal as accurately as possible. If we subtract the maximum-likelihood (best-fit waveform) from the data, we get:
\begin{equation}
    d_{\mathrm{sub}}(t) = n(t) + h_{\mathrm{mem}}(t) + \delta h_{\mathrm{cbc}}(t) ,
\end{equation}
Where $\delta h_{cbc}(t)$ is the difference between the true waveform (which we will never truly know) and our best approximation to it.

Greater the fidelity of the waveform model used, better is the subtraction, and smaller is $\delta h_{cbc}(t)$.
As shown in Fig. \ref{fig:cbc_vs_memory}, although the actual memory signal is a step-like function in time domain, GW detectors are not sensitive to DC offsets. Moreover, detector sensitivity drops drastically at lower frequencies. Thus, the memory signal observable in the detector bandwidth is not the actual step-like function, but rather a transient without any DC component and low-frequency components strongly suppressed by the detector's PSD. This does convert the steady-state memory signal into a short-duration transient, but this transient has a morphology quite distinct from that of the astrophysical compact binary merger signal.

Thus, the first step in the extraction of memory information from the residual data is to apply a highpass filter on it, which automatically highpasses the memory signal buried in the signal and mimics the memory signal that will be present in the detector's signal. The solid black waveform in Fig. \ref{fig:cbc_vs_memory} shows the memory waveform after highpassing. As the highpass filter is a linear operator, the amplitude of the highpassed memory transient is proportional to the DC memory offset $A$. The optimal SNR of the highpassed transient also is proportional to the memory offset. This opens up an opportunity of indirectly measuring the memory offset from the SNR of the memory signal. We can construct a {\it dictionary} which maps a range of memory offset values to optimal SNR values and use it to estimate memory offset, and hence the memory ratio from the highpassed memory alone.

Assuming that we know $\tau$, the rise time of the sigmoid memory signal, we can create a normalized template for the memory waveform - $\hat{h}$ with unit norm as $(\hat{h}|\hat{h}) = 1$, that is, 

\begin{equation}{\label{eqn:norm_mem_template}}
   4 \,\, \mathrm{Re} \bigg[ \int_0^{\infty} df \frac{|\hat{h}(f)|^2}{S_n(f)} \bigg] = 1 
\end{equation}

where $S_n(f)$ is the PSD of the detector. 
The filter is implemented through the \texttt{pycbc.filter.resample.highpass} module, which uses a time-domain Infinite Impulse Response (IIR) Butterworth filter of order = 8.
For CE and ET, the lower frequency cut for the SNR integral is 5Hz while for LIGO-India it is 10Hz. The upper cut 2048 Hz is half the sampling frequency 4096 Hz. 

Now, if the optimal matched-filter value of the memory signal is $a$, the highpassed memory signal can be written as:
\begin{equation}
    h_{\mathrm{mem}} = a \hat{h}
\end{equation}

Thus, the residual becomes:
\begin{equation}
    d_{\mathrm{substracted}} = n(t) + a \hat{h}(t) + \delta h_{cbc}(t)
\end{equation}

Performing matched-filter on both sides with a normalized memory template $\hat{h}$,
\begin{equation}
(d|\hat{h}) = (n|\hat{h}) + a + (\delta h_{cbc}|\hat{h})
\end{equation}

As the highpassed memory transient has a fundamentally different time-frequency morphology than a CBC chirp, the cross-term $(\delta h_{cbc}|\hat{h})$ is highly suppressed, reducing its contribution to the sum above. Thus, even if $\delta h_{cbc}(t)$ is not negligibly small to begin with, the inner product with the memory template serves to mitigate its effect. For the high-SNR events considered in this work, this leakage projection is sub-dominant to the Gaussian noise variance. 
As mentioned in the letter briefly, when the detector noise is Gaussian and stationary (we are ignoring detector glitches in this work), $(n|\hat{h})$, follows a standard normal distribution \citep{Maggiore:2007ulw}. Thus, $(d|\hat{h}) = (n|\hat{h}) + a$ is a shifted standard normal distribution, having a mean of $a$ and unity variance. Thus, for a given $A$, the output of the matched-filter statistic lies within a Gaussian envelope around $a$. As $A$ increases, $a$ increases in proportion. From a measured value of the matched-filter statistic (which will deviate from the optimal SNR owing to the presence of noise, but always within the allowed envelope), we can estimate the memory offset ${\hat A}$, and the Gaussian envelope around the $a$-vs-$A$ curve provides the error in $\hat{A}$ due to variation in noise. We apply this method to each detector individually, and extract their respective estimated ${\hat A}$ values. Fig. \ref{fig:bookkeeping} shows the \textit{book-keeping} plots used in our method. Once the individual error bounds on ${\hat A}$ are extracted for each detector in the network, the underlying polarization offsets (${\hat A}_+$ and ${\hat A}_\times$) are jointly reconstructed using the noise-weighted pseudoinverse of the antenna pattern matrix, as detailed in the main text.

\section{Error Analysis}

The method proposed in this work uses a combination of Bayesian and frequentist techniques applied in stages, and hence the final error in the estimates of the $B$ parameter contain contributions from different stages. The CBC parameters are estimated in the Bayesian framework, and the maximum likelihood waveform is subtracted from the data. As the posteriors are sharply localized and provide highly accurate estimates of the CBC parameters, the effect of widths of the posteriors on the study is minimal. In fact, if instead of the maximum likelihood waveform, we subtract the other waveforms from the data, the final results change by less than $1\%$. After the CBC signal has been dealt with, we turn to matched-filtering for estimating the memory offset. The memory signal being very weak (SNR $\sim 20$), and there being an inherent uncertainty in the matched-filter SNR due to the detector noise, the uncertainties in the memory ratio estimates for a single event are around $15-20\%$. From $\mathcal{R}$ values, we obtain the $B$ values not by inverting the (highly non-linear) relation between $B$ and $\mathcal{R}$, but rather by estimating the range of allowed $B$ values that are consistent with the observed $\mathcal{R}$ and $d_L$ values. To do this, we use Fig. 2 in the Letter as a book-keeping plot. Thus we cannot propagate the error analytically from $\mathcal{R}$ to $B$. But, given that the $\mathcal{R}$ uncertainties are clearly more than an order of magnitude above the CBC parameter uncertainties, it is safe to say that the major contributor to the observed $B$ uncertainty is the $\mathcal{R}$ uncertainty. Stacking multiple memory events can serve to reduce the uncertainty in the memory ratio estimates without increasing the uncertainty in the CBC parameter estimates for individual events. We anticipate this to have a positive impact on the reduction of the $B$ uncertainty - an idea we will investigate in detail in subsequent works. 

\bibliographystyle{aasjournalv7.1}
\bibliography{Final}{}



\end{document}